\documentclass[runningheads]{llncs}
\usepackage{graphicx}
\usepackage{makecell}
\usepackage{multirow}
\usepackage{amsfonts}   %
\usepackage[caption=false,font=footnotesize]{subfig}

\usepackage{amsmath}
\usepackage{xcolor}
\usepackage{hyperref}       %
\usepackage{url}            %
\definecolor{mypink1}{rgb}{0.858, 0.188, 0.478}

\begin{document}
\title{FocusLiteNN: High Efficiency Focus Quality Assessment for Digital Pathology}
\author{Zhongling Wang\inst{1} \and
Mahdi S. Hosseini\inst{2}\inst{3} \and
Adyn Miles\inst{2}  \and
Konstantinos N. Plataniotis\inst{2}  \and
Zhou Wang\inst{1} }

\authorrunning{Preprint. Accepted in MICCAI 2020.}
\institute{University of Waterloo, Waterloo, ON N2L 3G1, Canada \and
University of Toronto, Toronto, ON M5S 1A1, Canada \and
Huron Digital Pathology, St. Jacobs, ON, N0B 2N0, Canada \\
\email{\{zhongling.wang,zhou.wang\}@uwaterloo.ca}\\
\texttt{\{mahdi.hosseini,adyn.miles\}@mail.utoronto.ca,~kostas@ece.utoronto.ca}\\
\textcolor{mypink1}{\url{https://github.com/icbcbicc/FocusLiteNN}}}
\maketitle              %
\begin{abstract}
Out-of-focus microscopy lens in digital pathology is a critical bottleneck in high-throughput Whole Slide Image (WSI) scanning platforms, for which pixel-level automated Focus Quality Assessment (FQA) methods are highly desirable to help significantly accelerate the clinical workflows. Existing FQA methods include both knowledge-driven and data-driven approaches. While data-driven approaches such as Convolutional Neural Network (CNN) based methods have shown great promises, they are difficult to use in practice due to their high computational complexity and lack of transferability. Here, we propose a highly efficient CNN-based model that maintains fast computations similar to the knowledge-driven methods without excessive hardware requirements such as GPUs. We create a training dataset using FocusPath which encompasses diverse tissue slides across nine different stain colors, where the stain diversity greatly helps the model to learn diverse color spectrum and tissue structures. In our attempt to reduce the CNN complexity, we find with surprise that even trimming down the CNN to the minimal level, it still achieves a highly competitive performance. We introduce a novel comprehensive evaluation dataset, the largest of its kind, annotated and compiled from TCGA repository for model assessment and comparison, for which the proposed method exhibits superior precision-speed trade-off when compared with existing knowledge-driven and data-driven FQA approaches.

\end{abstract}
\section{Introduction}
The problem of out-of-focus microscopy lens in digital pathology is a huge bottleneck in existing high throughput Whole Slide Image (WSI) scanning platforms, making them difficult to be integrated in clinical workflows. WSI scans (aka digital slides) are required to be manually inspected for Focus Quality Assessment (FQA) on the pixel-level, which is (a) highly tedious and time consuming; and (b) subjective to an individual scoring that often causes inter/intra-variability issues. Both knowledge-driven and data-driven approaches have been developed to automate this process.

\textbf{Data-driven FQAs.} Recent developments involve supervised training of CNNs on the image patch labels of a given focus dataset of WSIs, where the network is either adopted from a pre-designed architecture followed by some minor adjustments \cite{campanella2018towards,kohlberger2019whole} or tailored from the scratch \cite{yang2018assessing,senaras2018deepfocus,pinkard2019deep}. The selection of training dataset can also be divided into two categories of either synthetically generating out-of-focus (defocus) images by convolving in-focus patches with artificial blur kernel with different grades (i.e. classes)  \cite{yang2018assessing,campanella2018towards,kohlberger2019whole}, or scanning tissue slides in different focal planes (z-levels) to generate real blur classes \cite{senaras2018deepfocus}. Existing open source software solutions such as CellProfiler \cite{mcquin2018cellprofiler} and HistoQC \cite{janowczyk2019histoqc} adopt variants of such models for FQA of WSIs. The high computational complexity and the lack of transferability are the main drawbacks of these models. 

\textbf{Knowledge-based FQAs.} Numerous methods have been developed in the literature based on a wide variety of domain knowledge, including human visual system models \cite{syntheticmaxpol,hosseini2019encoding}, microscopic optics models \cite{fqpath}, signal processing models \cite{gpc,lpc}, and natural image statistics models \cite{mlv,sparish}. For more information, please refer to \cite{fqpath} and references therein. Although these methods may have low computational cost, their precision performances are relatively low compared to data-driven solutions, as will be shown later.

\textbf{How Existing Models are Limited?} Despite great performances of data-driven approaches such as Convolutional Neural Network (CNN) in deep learning \cite{campanella2018towards,kohlberger2019whole,campanella2018towards,kohlberger2019whole}, they have not been integrated into high throughput scanners for QC control purposes due to two main reasons. First, the computational complexity of data-driven solutions is often too high to process GigaByte WSIs. We explain this in example as follows. Despite the FQA models take few seconds to process a patch from WSI that are fast enough, the story is quite different for high-throughput scanning systems. Depending on the vendors, several hundreds of glass slides can be mount in scanners (e.g. Philips Ultra Fast Scanner accepts 300 slides of 1''x3'' and Huron TissueScope-iQ accepts 400). In clinical settings, all scans should be completed during the night hours (less than 12 hours time frame) to be ready for diagnosis for the next day. Each slide is usually scanned at $0.5$um/pixel @$20$X magnification, containing $\sim{1}\text{cm}\times{1}\text{cm}$ tissue which translates to $25,000\times{25,000}$ digital WSI, yielding $\sim 2,500$ patches of $1024\times 1024$ ($50\%$ overlap). Assuming two models are used for assessment, i.e. M1: DenseNet-10 and M2: FocusLiteNN (our proposed model), the time taken for two models to complete the task is
\begin{align}
M1:~2,500\text{(patches/WSI)}\times{300}\text{(WSI)}\times 0.355\text{sec/patch} = 73.96~ \text{hour} \nonumber\\
M2:~2,500\text{(patches/WSI)}\times{300}\text{(WSI)}\times 0.017\text{sec/patch} = 3.54~\text{hour} \nonumber
\end{align}
Clearly, the speed gain from model M2 over M1 is obvious. The limitation in computational resources becomes equally important as the precision when choosing FQA models for GigaByte WSI processing \cite{gupta2019deep,topol2019high}. The second limitation is the lack of transferability of CNNs which becomes a barrier to process WSIs across different tissue stains and scanner variations. 

\textbf{Contributions.} Our aim in this paper is to address the challenges in data-driven FQAs. In particular, (a) we build a highly-efficient extremely light weight CNN-based model i.e. \textcolor{mypink1}{FocusLiteNN}\footnote{Codes and  models are available at \textcolor{mypink1}{\url{https://github.com/icbcbicc/FocusLiteNN}}} that maintains fast computations similar to the knowledge driven methods without excessive hardware requirements such as GPUs. The database used for training plays a crucial role, for which we suggest a training dataset using FocusPath \cite{hosseini2019encoding} which encompasses diverse tissue slides across nine different stain colors. We hypothesise that the stain diversity greatly helps the model to learn diverse color spectrum and tissue structures. (b) For algorithm evaluation and comparison, we introduce a novel comprehensive evaluation dataset that is annotated and compiled from TCGA repository. Comprehensive experiments and analysis are conducted that demonstrate the superior precision-speed compromise of the proposed approach.

\section{FocusLiteNN: Extremely Light-Weight CNN for FQA}
The main idea of our model design here is to reduce the layer complication in deep learning for FQA in digital pathology, while still being able to benefit from machine learning framework to adapt data distribution for generalization. We build a simple data-driven model, called FocusLiteNN, which includes only one convolution layer (very shallow) for feature transformation. Such shallow design is based on the assumption that out-of-focus blur in digital pathology can be characterized using a relatively simple model since (a) the distortion process is taken place in a well controlled environment (within the WSI scanner); and (b) focus information is mainly encoded in the low-level (edge) information rather than high-level (semantic) information.

Let us assume that the sharpness level is uniform within a small enough mosaic patch $\mathbf{X} \in \mathbb{R}^{H \times W \times 3}$ extracted from a WSI scan. The idea is to first convolve the image patch with a kernel set $\mathbf{\Phi}\in\mathbb{R}^{{h}\times{w}\times{3}\times{N}}$ and then apply a non-linear pooling function to predict the sharpness of the input patch
\begin{equation} 
y = p_{\text{NL}}\left(\sum^{3}_{k=1}\mathbf{\Phi}_k\ast\mathbf{X_k} + \mathbf{b}\right)
\label{LW_CNN}
\end{equation}
where, $\mathbf{\Phi}_{k} \in \mathbb{R}^{{h}\times{w}\times{N}}$ is the convolution kernel for $k$th input channel. Here, $\mathbf{X_k} \in \mathbb{R}^{{H}\times{W}}$ is $k$th channel of input patch and $\mathbf{b}\in \mathbb{R}^N$ is a bias vector. $y \in \mathbb{R}$ is the predicted score of $\mathbf{X}$. The 2D convolution operator $\ast$ is applied with a stride of $5$ and $p_{\text{NL}}$ is a non-linear pooling function which maps a $2$D response to an overall sharpness score $y \in \mathbb{R}$. We set the kernel size to $h=w=7$ for all experiments. The use of pooling function $p_{\text{NL}}$ is also critical: by adding non-linearity to the model, it greatly enhances the approximation capability of the simple model. We defined it as
\begin{equation}
p_{\text{NL}}(\mathbf{x}) = \mathbf{w_1} \cdot \min(\mathbf{x}) + \mathbf{w_2} \cdot \max(\mathbf{x}) + w_3
\end{equation}
where, $\mathbf{x} \in \mathbb{R}^{\frac{H-h+7}{5} \times \frac{W-w+7}{5} \times N}$ are the responses produced by convolution and $\mathbf{w_1} \in \mathbb{R}^N, \mathbf{w_2} \in \mathbb{R}^N, w_3 \in \mathbb{R}$ are trainable parameters. The use of channel-wise 2D max and min in $p_{\text{NL}}$ makes the model capable of capturing extreme kernel responses. We refer the model in (\ref{LW_CNN}) as $N$-kernel mode of FocusLiteNN.
\begin{figure}[htp]
    \centering
     \subfloat[][\centering{\tiny{$\phi(x_1,x_2)$}}]{\includegraphics[height=0.3\textwidth]{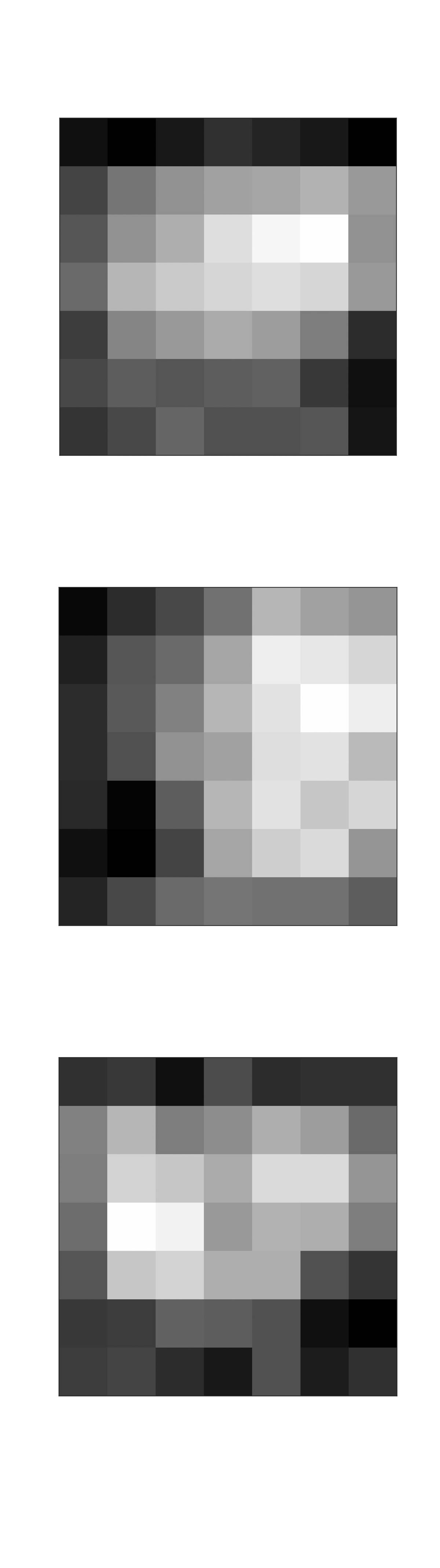}}
     \subfloat[][\centering{\tiny{$|\hat{\phi}(\omega_1,\omega_2)|$}}]{\includegraphics[height=0.3\textwidth]{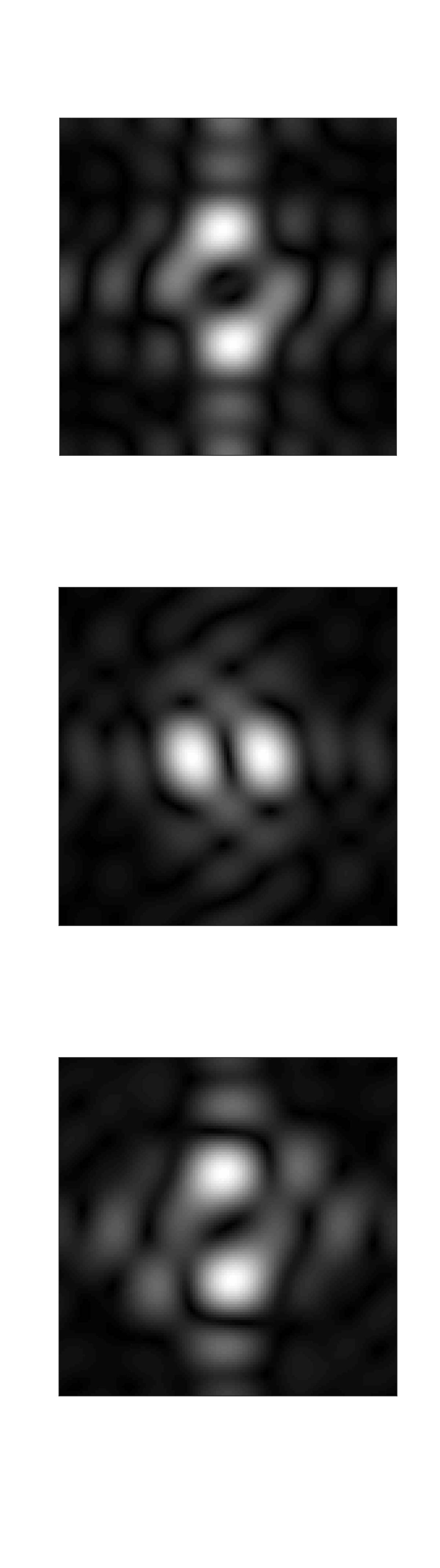}}
     \subfloat[][\centering{\tiny{$\angle\hat{\phi}(\omega_1,\omega_2)$}}]{\includegraphics[height=0.3\textwidth]{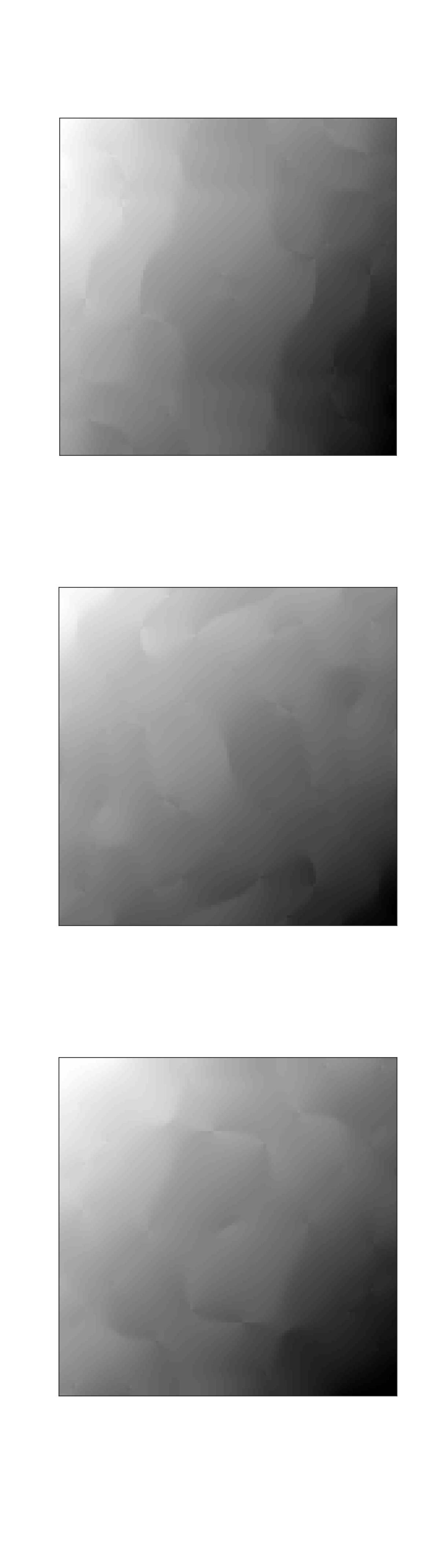}}
     \subfloat[][\centering{\tiny{cross-sections: vert/horz resp.}}]{\includegraphics[height=0.3\textwidth]{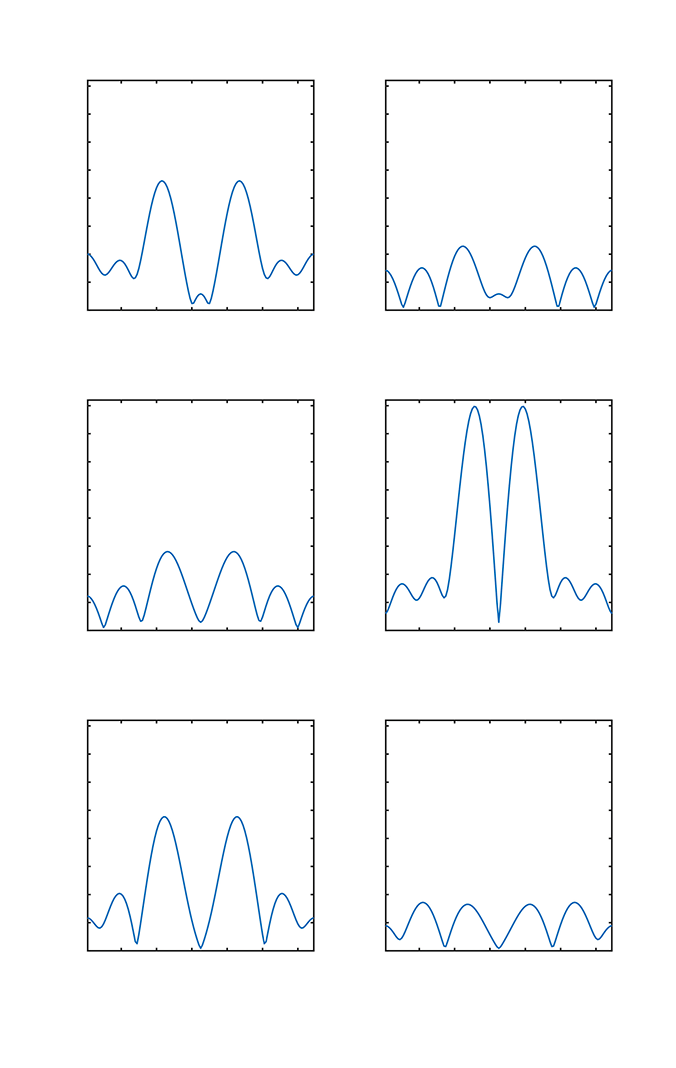}}
    \caption{Filter responses of 1-kernel mode FocusLiteNN, shown for (a) spatial representation, (b) frequency amplitude, (c) unwrapped frequency phase; and (d) cross-sections (vertical/horizontal) for frequency amplitudes.}
    \label{LW_CNN_kernel_responses}
\end{figure}
Here we demonstrate the kernel response in FocusLiteNN (1-kernel) for each color channel in Figure \ref{LW_CNN_kernel_responses}. The filter responses are shown for both spatial representation (aka impulse response) i.e. $\phi(x_1,x_2)$, magnitude frequency response $|\hat{\phi}(\omega_1,\omega_2)|$ and phase frequency response $\angle\hat{\phi}(\omega_1,\omega_2)$. Note that the filter responses are mainly significant along the perpendicular axes (i.e. horizontal and vertical) rather than rotational angles.

\section{Selection of Dataset}
The development of data-driven FQA in digital pathology heavily relies on the selection of dataset for training. While the CNN models perform very well on training dataset, the ultimate question is how well the models can be transferred to other dataset for evaluation? This is of paramount importance in digital pathology where the models should be capable of (a) accurately predicting focus scores on the slides regardless of tissue structures and staining protocols; and (b) accounting for color disparities that could cause by WSI scanner variations and tissue preparation in different pathology labs.

\subsection{FocusPath for CNN Training}
\begin{figure}[htp]
    \centering
    \subfloat[][\centering{\tiny{Stain dist. in FocusPath}}]{\includegraphics[height=0.15\textwidth]{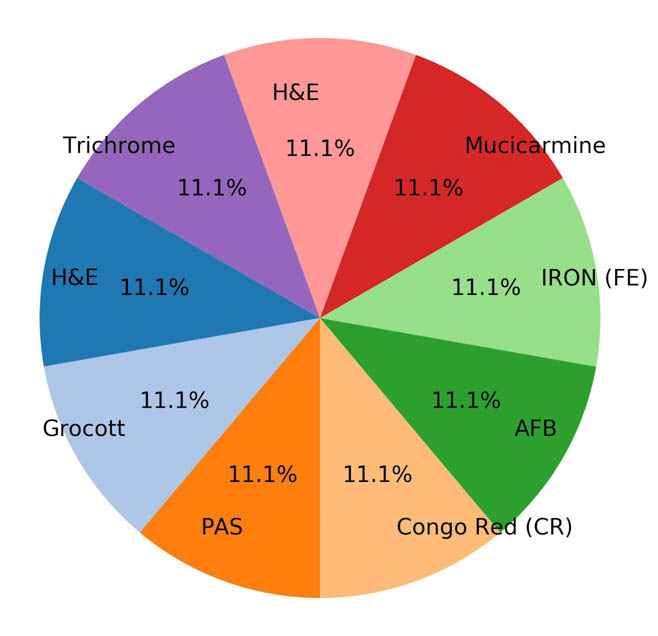}\label{FocusPath_Stat_Dist}}
    \subfloat[][\centering{\tiny{Different ground-truth focus levels in FocusPath  }}]{\includegraphics[height=0.15\textwidth]{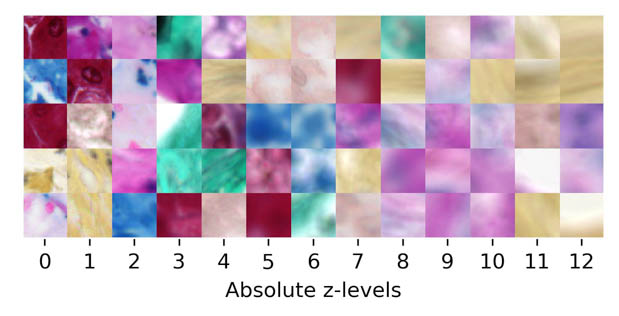}\label{FocusPath_patch_examples}}
    \subfloat[][\centering{\tiny{Pie Chart of organ dist. for TCGA@Focus}}]{\includegraphics[height=0.25\textwidth]{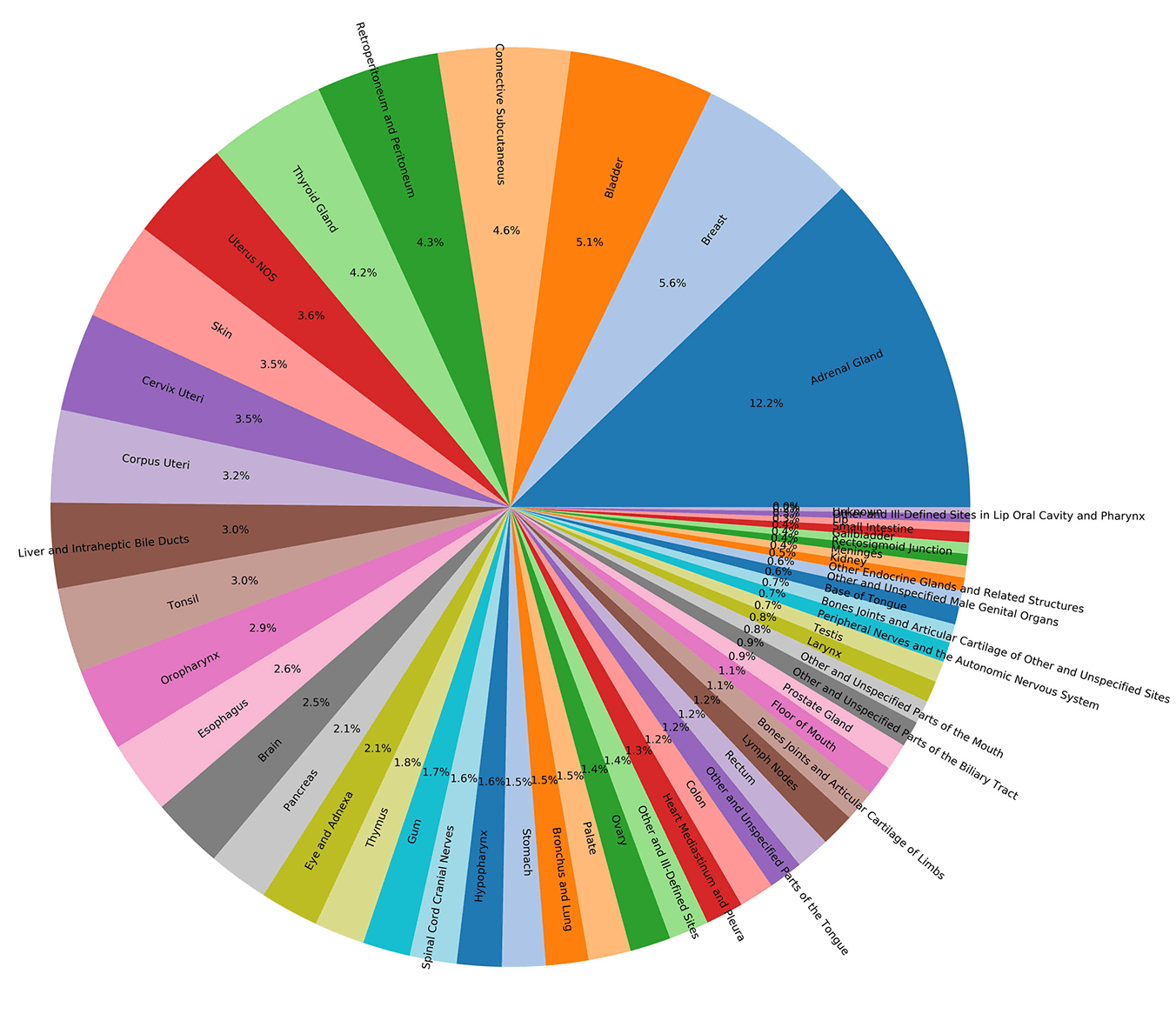}\label{TCGA_Stat_Dist}}
    \subfloat[][\centering{\tiny{In- and Out- focus examples in TCGA@Focus}}]{\includegraphics[height=0.225\textwidth]{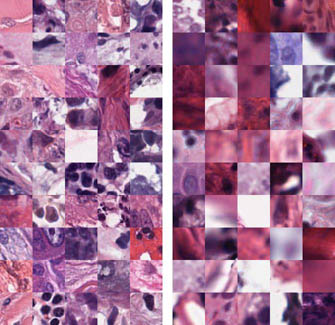}\label{TCGA_patch_examples}}
    \caption{(a)-(b) FocusPath dataset \cite{hosseini2019encoding} containing $8,640$ image patches distributed among nine different tissue stains and annotated in $15$ different focus levels. (c)-(d) The TCGA@Focus dataset containing $14,371$ image patches compiled with both in- and out- focus labels.}
\end{figure}

The \textcolor{mypink1}{FocusPath}\footnote{The data is available at \textcolor{mypink1}{\url{https://zenodo.org/record/3926181}}} dataset \cite{hosseini2019encoding} contains $8,640$ patches of $1024\times 1024$ image extracted from nine different stained slides. The WSIs are scanned by Huron TissueScope LE1.2 \cite{dixon2014pathology} using $40$X optics lens @$0.25\mu$m/pixel resolution. Each patch is associated with different focal plane (i.e. absolute z-level score) ranging from $\{0,\cdots, 14\}$ corresponding to the ground-truth class for focus level. The statistical distribution of color stains in FocusPath is shown in Figure \ref{FocusPath_Stat_Dist} and the patch examples are shown in Figure \ref{FocusPath_patch_examples} for different variations of focus levels.

Since the FocusPath includes diverse color stains compiled with different tissue structures, this makes the dataset well suited for developments of data-driven FQA models. Furthermore, we hypothesize that the diversity of color stains greatly helps generalize the CNN training to different tissues structures and color spectrum--no color augmentation is required such as in \cite{stacke2019closer}.
\subsection{TCGA@Focus--An Evaluation Benchmark}
A dataset of $1000$ WSIs was selected from The Cancer Genome Atlas (TCGA) repository in SVS format gathered from $52$ organ types provided by the National Cancer Institute (NCI) / National Institutes of Health (NIH) \cite{tomczak2015cancer}. The statistical distribution of the number of slides per organ site is shown in Figure \ref{TCGA_Stat_Dist}. Note that the diversity of the organ types here is important to include wide spectrum of tissue textures and color information caused by variations in staining and WSI scans. Our goal here was to annotate two different categories of \textit{``in-focus''} and \textit{``out-focus''} regions-of-interests (ROI) within each slide corresponding to the binary ground truth scores of ``$1$'' and ``$0$'', respectively. The patch examples of each category are shown in Figure \ref{TCGA_patch_examples} for different organ types. The compiled dataset is called \textcolor{mypink1}{TCGA@Focus}\footnote{The data is available at \textcolor{mypink1}{\url{https://zenodo.org/record/3910757}}} and contains $14,371$ image patches in total, where $11,328$ patches are labeled in-focus and $3,043$ patches out-focus.

\section{Experiments}
~~~\textbf{Model Selection and Evaluation.} We adopt five different categories in knowledge based methods for the experiments using (1) human visual system: Synthetic-MaxPol \cite{syntheticmaxpol}, and HVS-MaxPol-1/HVS-MaxPol-2 \cite{hosseini2019encoding}, (2) microscopy lens modeling: FQPath \cite{fqpath}, (3) natural image statistics: MLV \cite{mlv}, SPARISH \cite{sparish}; and (4) signal processing based: GPC \cite{gpc} and LPC \cite{lpc}. For data-driven methods we select a diverse range of CNN models  in terms architecture complexity using EONSS \cite{eonss} with four conv layers developed for the purpose of Image Quality Assessment (IQA), as well as DenseNet-13 \cite{densenet} (eight conv layers) and variations of ResNet \cite{resnet} ($8$, $48$, and $99$ conv layers) developed for computer vision applications. We evaluate selected FQA models in terms of statistical correlation and classification performance as well as computational complexity on the FocusPath and TCGA@Focus datasets. At the end, we also show the heat maps generated by these models on a sample image.

\textbf{Implementation Details.} All CNNs are re-trained on the FocusPath dataset with the same pre-processing techniques, optimizer and loss function. The FocusPath dataset is randomly split into a train ($60\%$) - validation ($20\%$) - test ($20\%$). The validation subset is used to determine the hyper-parameters. Training and testing are repeated in $10$ folds of splits and the average performance is reported. All models are transferred to TCGA@Focus dataset for evaluation. The input dimensions for all CNNs are set to $235 \times 235 \times 3$. During testing, we densely sample the original patches with a stride of $128 \times 128$ and the average score is taken as the overall sharpness. Adam optimizer is utilized for all models. For FocusLiteNN, the learning rate is set to 0.01 with decay interval of 60 epochs. For other models, the learning rate is set to 0.001 with decay interval of 40 epochs. Each model is trained for 120 epochs to ensure convergence. The Pearson Linear Correlation Coefficient (PLCC) is used as the loss function for all models. PLCC bounds the loss value between -1 and 1, which helps to stabilize the training process.

\begin{figure}[h!]
    \centering
    \captionsetup[subfloat]{position=bottom,labelformat=empty}
    \subfloat{\includegraphics[width=0.2\textwidth]{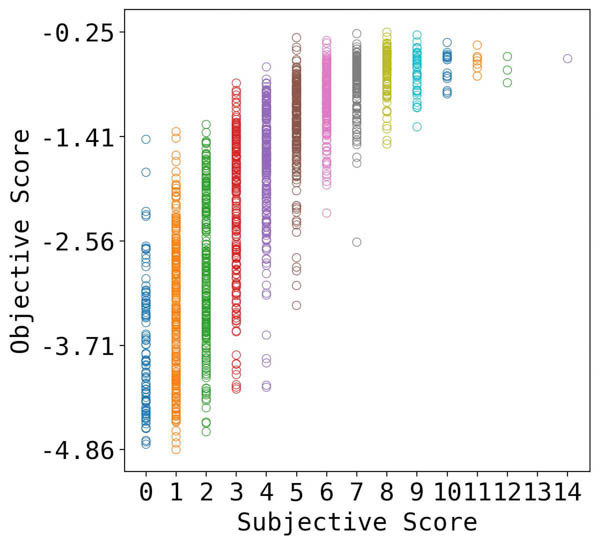}}
    \subfloat{\includegraphics[width=0.2\textwidth]{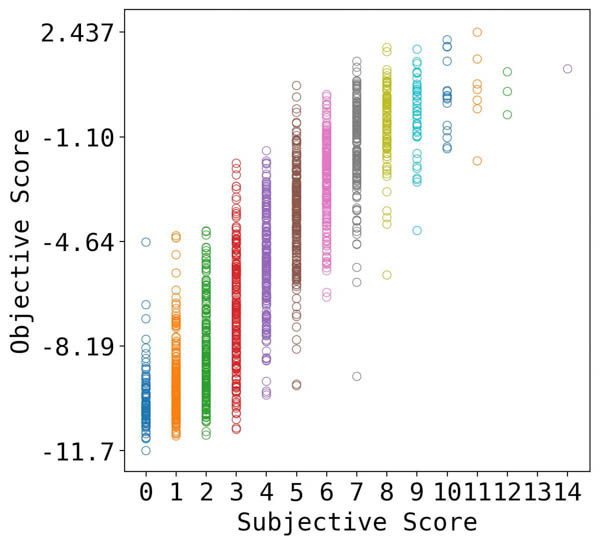}}
    \subfloat{\includegraphics[width=0.2\textwidth]{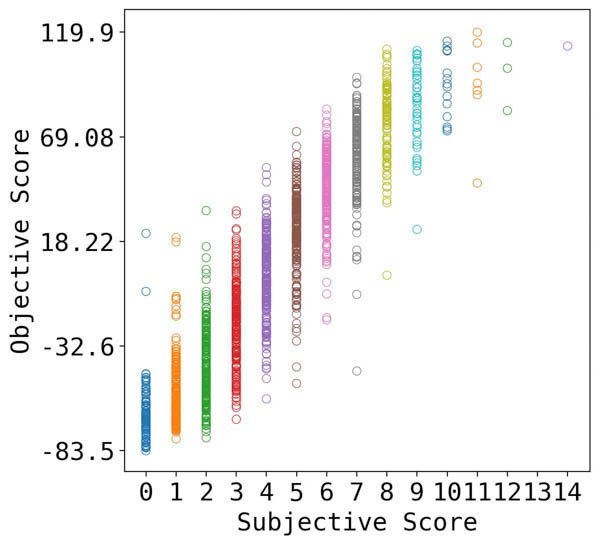}}
    \subfloat{\includegraphics[width=0.2\textwidth]{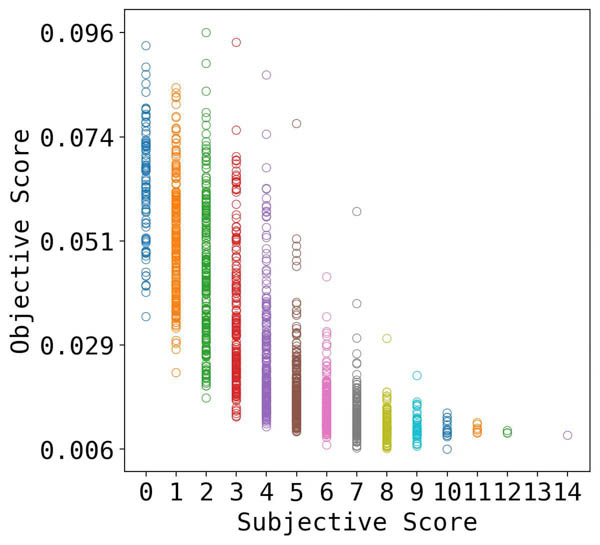}}\\
    \subfloat{\includegraphics[width=0.2\textwidth]{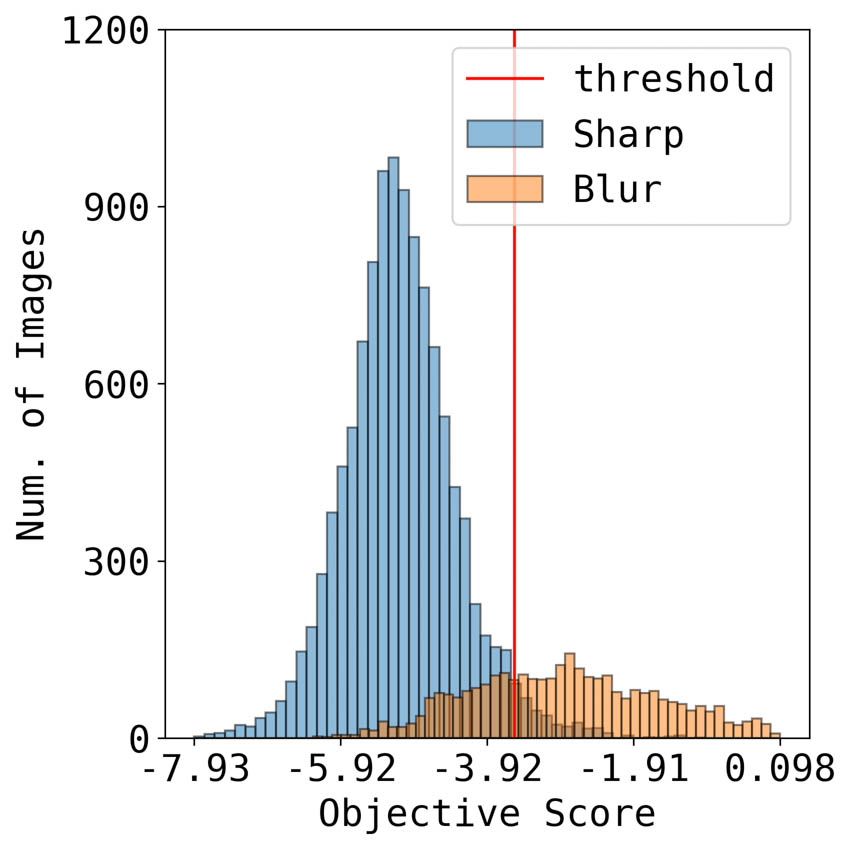}}
    \subfloat{\includegraphics[width=0.2\textwidth]{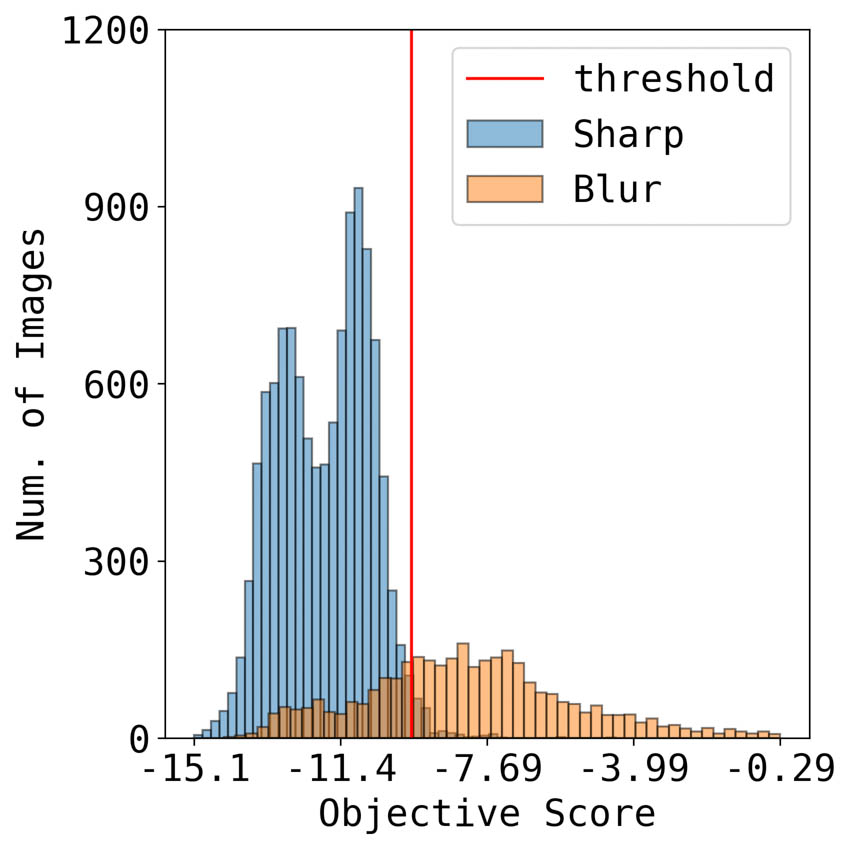}}
    \subfloat{\includegraphics[width=0.2\textwidth]{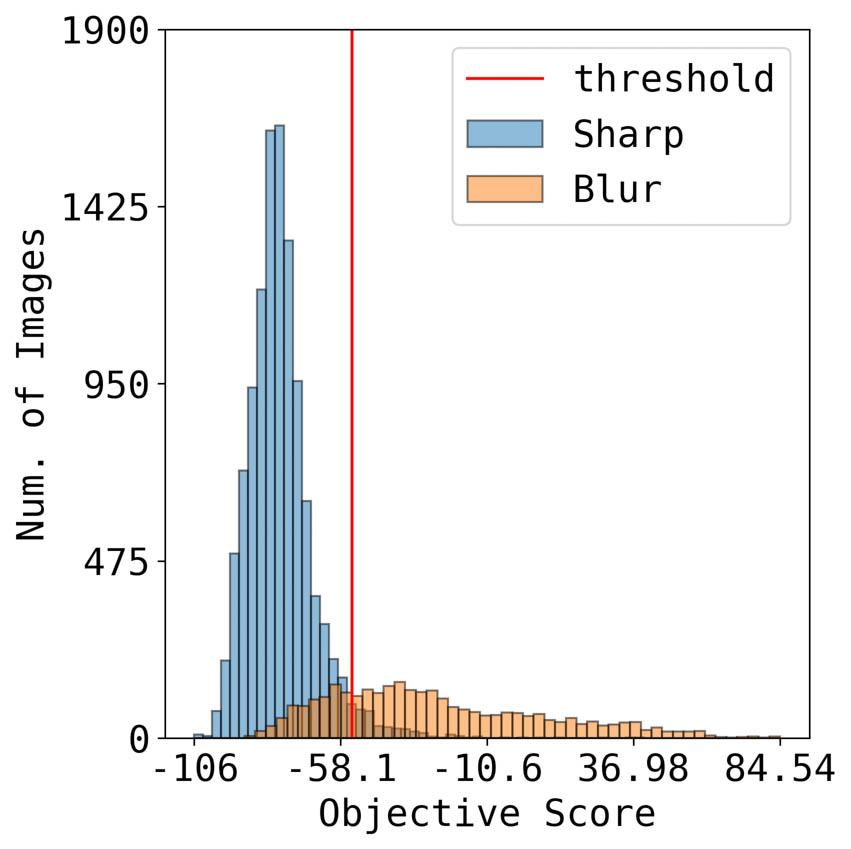}}
    \subfloat{\includegraphics[width=0.2\textwidth]{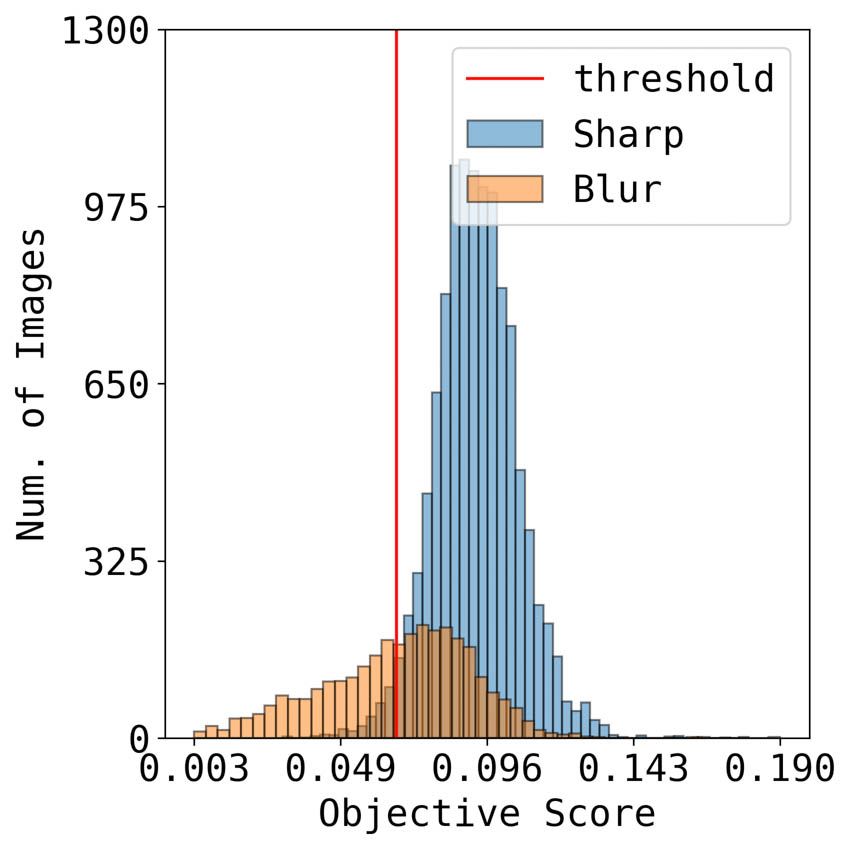}}\\\hspace{.2in}
    \subfloat[][\centering{\scriptsize{FocusLiteNN (1-kernel)}}]{\includegraphics[width=0.2\textwidth]{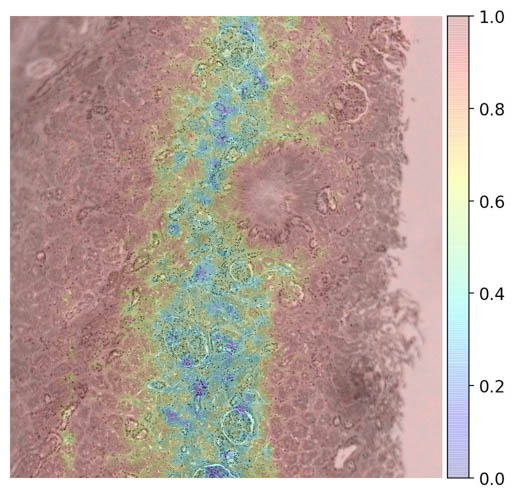}}
    \subfloat[][\centering{\scriptsize{EONSS \cite{eonss}}}]{\includegraphics[width=0.2\textwidth]{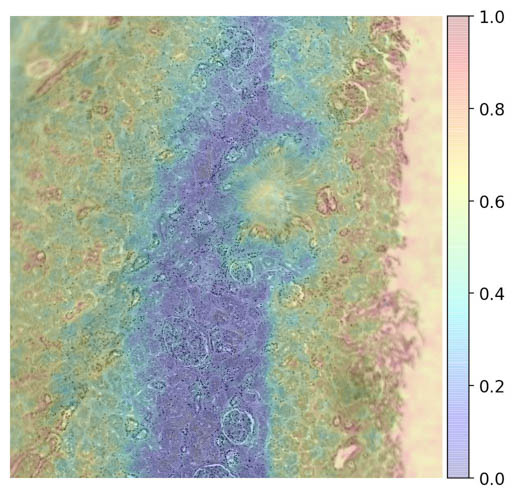}}
    \subfloat[][\centering{\scriptsize{ResNet-50 \cite{resnet}}}]{\includegraphics[width=0.2\textwidth]{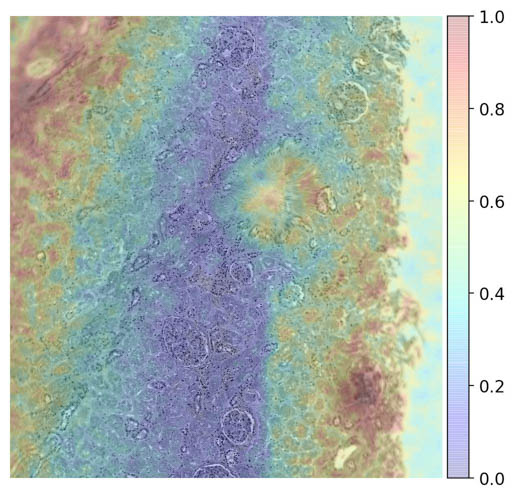}}
    \subfloat[][\centering{\scriptsize{MLV \cite{mlv}}}]{\includegraphics[width=0.2\textwidth]{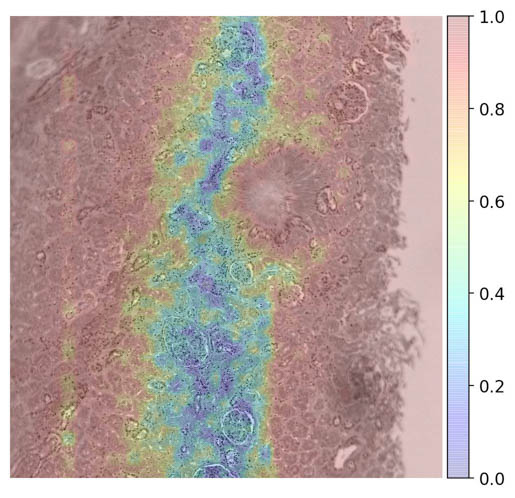}}\\
    \subfloat[][\centering{\scriptsize{ROC-AUC v.s. CPU Time}}]{\includegraphics[height=0.4\textwidth]{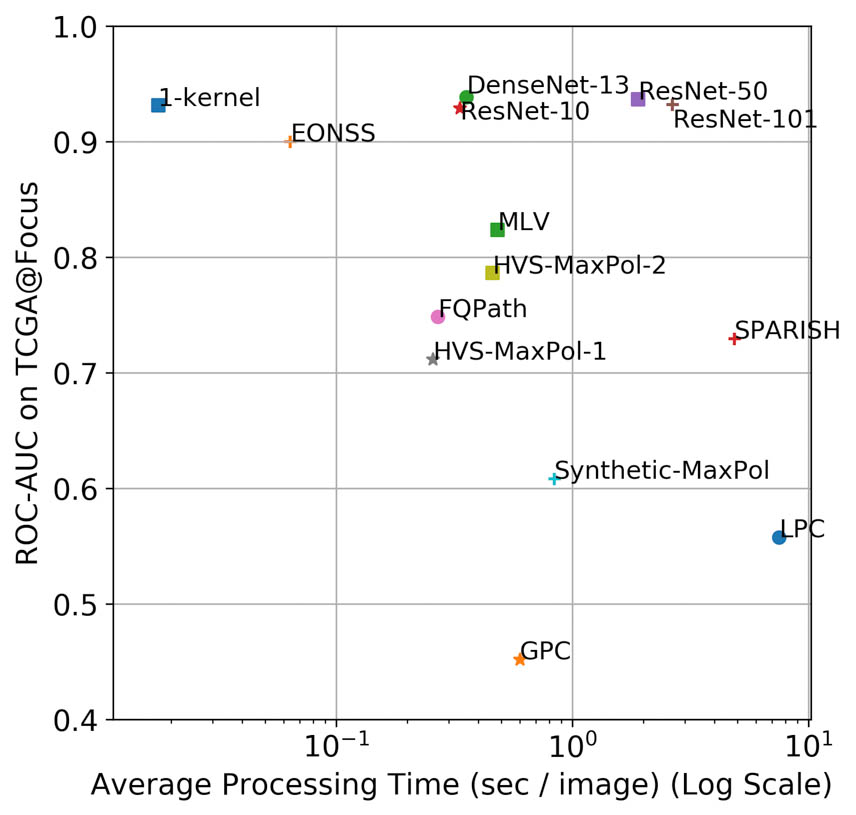}}
    \subfloat[][\centering{\scriptsize{ROC-AUC v.s. \# Model Params}}]{\includegraphics[height=0.4\textwidth]{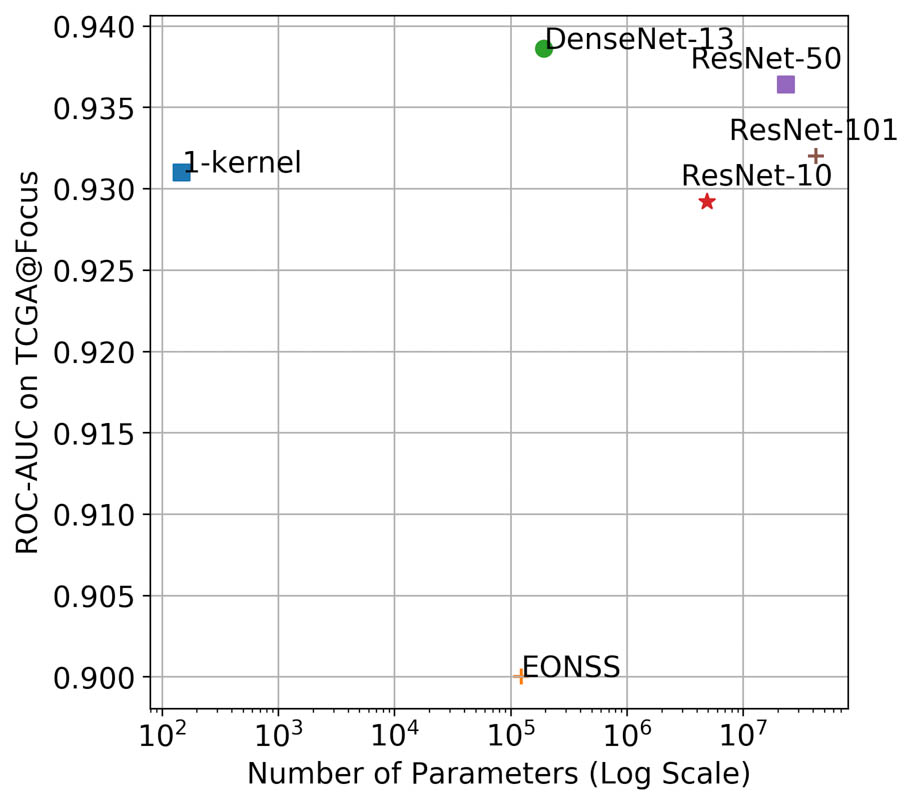}}
    \caption{Evaluation results for 4 models. \textbf{First row}: scatter plots of absolute z-level versus predicted scores on FocusPath dataset. \textbf{Second row}: histogram of objective scores on TCGA@Focus. \textbf{Third row}: normalized heat maps. Higher score indicate more blurriness. \textbf{Forth row}: average Processing Time versus ROC-AUC and model size versus ROC-AUC on TCGA@Focus Dataset. Please refer to the Supplementary Materials (Fig. 2) for the complete results of $16$ models.}
    \label{eval_fig}
\end{figure}

\textbf{Performance Evaluation.} The metrics used to evaluate the performances are Spearman's Rank Correlation Coefficient (SRCC), PLCC, Area Under the Curve of the Receiver Operating Characteristic curve ($\mathrm{ROC}$), Area Under the Curve of the Precision Recall Curve ($\mathrm{PR}$). SRCC measures the monotonicity between the predicted sharpness score and the absolute z-level, while PLCC measures the linear correlation between them. When measuring $\mathrm{ROC}$ and $\mathrm{PR}$ on the FocusPath dataset, we first binarize the z-levels by considering all patches with absolute z-level $0$, $1$, $2$ as sharp and those equal or larger than $2$ as blurry. The results are shown in Table \ref{table:all_results}.
\begin{table}[h!]
	\scriptsize
	\caption{SRCC, PLCC, ROC-AUC, PR-AUC Performance of 16 FQA Models on FocusPath Dataset and TCGA@Focus Dataset. }
	\label{table:all_results}
	\centering
	\begin{tabular}{|c|c||c|c|c|c||c|c||r|c|}
		\Xhline{2\arrayrulewidth}
		\multirow{2}{*}{Type} & \multirow{2}{*}{Model} & \multicolumn{4}{c||}{FocusPath} & \multicolumn{2}{c||}{TCGA@Focus} & \multirow{2}{*}{Size}  & Time \\
		\cline{3-8}
		& & SRCC & PLCC & $\mathrm{ROC}$ & $\mathrm{PR}$ & $\mathrm{ROC}$ & $\mathrm{PR}$ &  & (sec)\\ 
		\Xhline{2\arrayrulewidth}
		\multirow{8}{*}{\rotatebox[origin=c]{90}{Data-driven based}} 
		& FocusLiteNN (1-kernel)                               & 0.8766            & 0.8668            & 0.9468            & 0.9768            & 0.9310            & 0.8459            & \textbf{148}  & \textbf{0.017} \\
		& FocusLiteNN (2-kernel)                               & 0.8782            & 0.8686            & 0.9481            & 0.9770            & \textbf{0.9337}   & 0.8499            & \textbf{299}  & \textbf{0.019} \\
		& FocusLiteNN (10-kernel)                              & 0.8931            & 0.8857            & 0.9542            & 0.9802            & 0.9322            & \textbf{0.8510}   & \textbf{1.5K}& \textbf{0.019} \\
		& EONSS \cite{eonss}                                & 0.9009            & 0.8951            & 0.9540            & 0.9799            & 0.9000            & 0.8473            & 123K          & 0.063 \\
		& DenseNet-13 \cite{densenet}                       & \textbf{0.9253}   & \textbf{0.9197}   & \textbf{0.9662}   & 0.9849            & \textbf{0.9386}   & \textbf{0.8646}   & 193K          & 0.355 \\
		& ResNet-10 \cite{resnet}                           & \textbf{0.9278}   & \textbf{0.9232}   & \textbf{0.9671}   & \textbf{0.9853}   & 0.9292            & \textbf{0.8559}   & 4.9M        & 0.334 \\
		& ResNet-50 \cite{resnet}                           & \textbf{0.9286}   & \textbf{0.9244}   & \textbf{0.9676}   & \textbf{0.9855}   & \textbf{0.9364}   & 0.8144            & 24M       & 1.899 \\
		& ResNet-101 \cite{resnet}                          & 0.9242            & 0.9191            & 0.9644            & \textbf{0.9840}   & 0.9320            & 0.8447            & 43M       & 2.655 \\
		\hline
		\multirow{8}{*}{\rotatebox[origin=c]{90}{Knowledge based}} 
		& FQPath \cite{fqpath}                              & 0.8395            & 0.8295            & 0.9375            & 0.9739            & 0.7483            & 0.6274            & N.A.          & 0.269 \\
		& HVS-MaxPol-l \cite{hosseini2019encoding}          & 0.8044            & 0.8068            & 0.9400            & 0.9743            & 0.7118            & 0.5622            & N.A.          & 0.257 \\
		& HVS-MaxPol-2 \cite{hosseini2019encoding}          & 0.8418            & 0.8330            & 0.9434            & 0.9757            & 0.7861            & 0.6721            & N.A.          & 0.458 \\
		& Synthetic-MaxPol \cite{syntheticmaxpol}           & 0.8243            & 0.8139            & 0.9293            & 0.9707            & 0.6084            & 0.4617            & N.A.          & 0.841 \\
		& LPC \cite{lpc}                                    & 0.8375            & 0.8321            & 0.9223            & 0.9681            & 0.5576            & 0.4564            & N.A.          & 7.510 \\
		& GPC \cite{gpc}                                    & 0.7851            & 0.7602            & 0.9095            & 0.9604            & 0.4519            & 0.2830            & N.A.          & 0.599 \\
		& MLV \cite{mlv}                                    & 0.8623            & 0.8528            & 0.9414            & 0.9758            & 0.8235            & 0.6943            & N.A.          & 0.482 \\
		& SPARISH \cite{sparish}	                        & 0.3225            & 0.3398            & 0.7724            & 0.8875            & 0.7293            & 0.6414            & N.A.          & 4.853 \\
		\Xhline{2\arrayrulewidth}
	\end{tabular}
\end{table}
On the FocusPath dataset, the overall performance of DenseNet-13 \cite{densenet}, ResNet-10 \cite{resnet}, ResNet-50 \cite{resnet} and ResNet-101 \cite{resnet} in all 6 metrics are the best and are similar to each other. Assuming that the testing subset of FocusPath is drawn from the same distribution as of the training subset, this observation shows that those data-driven based models with more parameters can fit the distribution of training data better. ResNet-50, the best performer among deep CNN based models, outperforms the 10-kernel model, the best performer among shallow CNN based models, by $3.5\%$ in SRCC and $2\%$ in $\mathrm{ROC}$. To visualize the statistical correlation of all models, the scatter plots of the predicted scores versus z-levels on the FocusPath testing subset are shown in the first row of Fig \ref{eval_fig}. We can see that the monotonisity and linearity between the prediction and ground truth are best preserved in deep CNN base models.

All models are also evaluated on the TCGA@Focus dataset to study the trasnferability performance where no training is involved. Here, DenseNet-13 \cite{densenet} achieves the highest scores on both ROC-AUC and PR-AUC. While the overall performance of the deep CNN based models are still in the top tier, the gap between them and the shallow CNNs are getting smaller compared with the performance difference on FocusPath dataset: ResNet-50 only outperforms the FocusLiteNN (10-kernel) model by $0.4\%$ in terms of $\mathrm{ROC}$. Distribution of the predicted scores on the TCGA@Focus dataset and their ground truth labels as well as the classification thresholds for all models are also shown in the second and third rows of Fig \ref{eval_fig}.

\textbf{Computational Complexity Analysis.} The testing image is $1024 \times 1024 \times 3$ 8-bit in the FocusPath dataset. Two experiments are conducted, the first one is ROC-AUC on the TCGA@Focus dataset versus CPU time (Fig \ref{eval_fig} last row left). To fairly compare the computational complexity, all models are running on an Intel i9-7920X @ 2.90GHz with 32 GB memory. Image reading time is excluded from the CPU time, but the pre-processing time for each model, such as dense sampling, is measured. The MontCarlo simulation is done for $100$ times and the average is reported. The second experiment is ROC-AUC on the TCGA@Focus dataset versus number of model parameters (Fig \ref{eval_fig} last row right). We count the number of trainable parameters of the data-driven models and plot the numbers against their performance. We can clearly see that the 1-kernel model outperforms others by a large margin in terms of both CPU time and model size: it outperforms the second fast model EONSS \cite{eonss} by $3.4\%$ in terms of ROC-AUC, but consuming only $27\%$ of its CPU time with $0.1\%$ of its model size.

\textbf{Heat Map Visualization.} To better visualize the model outputs, we generate heat maps for each model, as shown in Fig \ref{eval_fig} (further heatmaps are provided in Supplementary Materials Fig 1). For all models, we densely sample $235 \times 235$ patches from the WSI scan with a stride of $128 \times 128$ for scoring and interpolated accordingly. These scores are then mapped to colors and overlaid on the grayscale version of the scan. The most blurry parts are in the upper left corner, lower right corner, and in the circle in the middle. The vertical strip taken up $\frac{1}{3}$ of the space is in focus. In Fig \ref{eval_fig} the third row, we showed the relative blurriness level within a scan by normalizing the scores to the range 0 to 1 before color mapping. Knowledge based models and FocusLiteNN prefer to predict the entire scan as more blurry even for in focus part. Deep CNN-based models such as EONSS \cite{eonss}, ResNet \cite{resnet} and DenseNet \cite{densenet} are less aggressive and can identify in focus regions, which are more perceptually accurate. To demonstrate the absolute blurriness level of a scan, we train the FocusLiteNN (1-kernel), ResNet-10 \cite{resnet} and EONSS \cite{eonss} with MSE loss on the FocusPath dataset. The predicted scores correspond to absolute z-levels in the FocusPath dataset. The results are shown in the the first row of Supplementary Materials Fig 1.

\section{Conclusion}
We propose a highly efficient CNN-based automated FQA approach, aiming to accelerate the clinical workflow in WSI scanning. Reaching the performance of more complex models with fewer parameters is one of our main contributions. We use FocusPath to create a training dataset with diverse tissue slides and stain colors, which helps the model to learn diverse color spectrum and tissue structures. We introduce a novel comprehensive evaluation dataset annotated and compiled from TCGA repository, the largest of its kind for assessing and comparing FQA models. Our test results show that the proposed model demonstrates superior precision-speed trade-off when compared with existing knowledge-driven and data-driven FQA approaches. A somewhat surprising finding in our study is that even when we trim down our CNN model to the minimal 1-kernel size, it still maintains a highly competitive performance and transferability. In conclusion, our proposed FocusLiteNN surpasses (by a large margin) all SOTA models in terms of speed, and yet achieves competitive performance with best accuracy model i.e. DenseNet-13.

\section*{acknowledgements}
The authors would like to greatly thank \textit{Huron Digital Pathology} (St. Jacobs, ON N0B 2N0, Canada) for the support of this research work.

\bibliographystyle{splncs04}
\bibliography{ref}

\section*{Appendix}
\section{Complete Figure Lists}
\begin{figure}[h!]
	\centering
	\subfloat[][\centering{\scriptsize{original}}]{\includegraphics[width=0.18\textwidth]{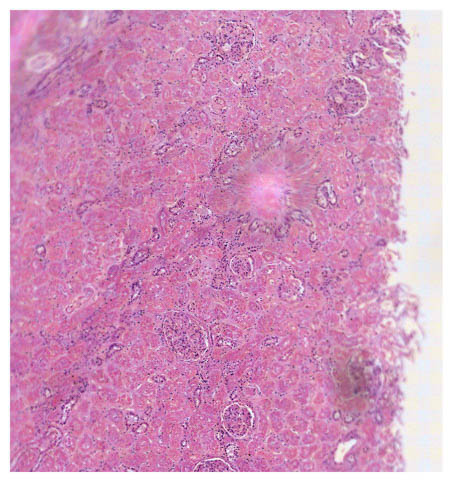}}
	\subfloat[][\centering{\scriptsize{1-kernel}}]{\includegraphics[width=0.2\textwidth]{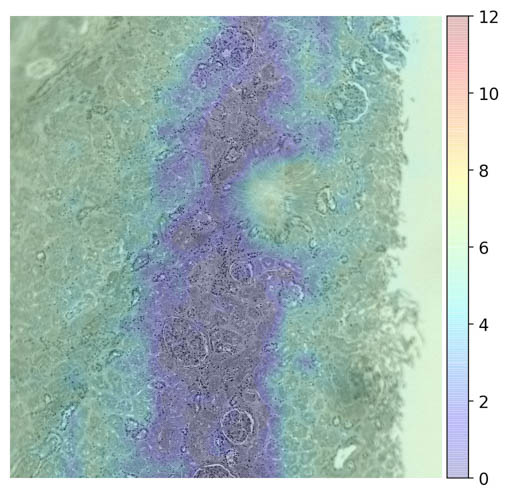}}
	\subfloat[][\centering{\scriptsize{ResNet-10}}]{\includegraphics[width=0.2\textwidth]{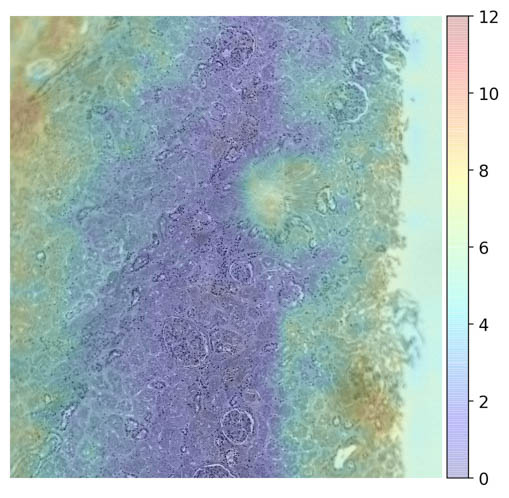}}
	\subfloat[][\centering{\scriptsize{EONSS}}]{\includegraphics[width=0.2\textwidth]{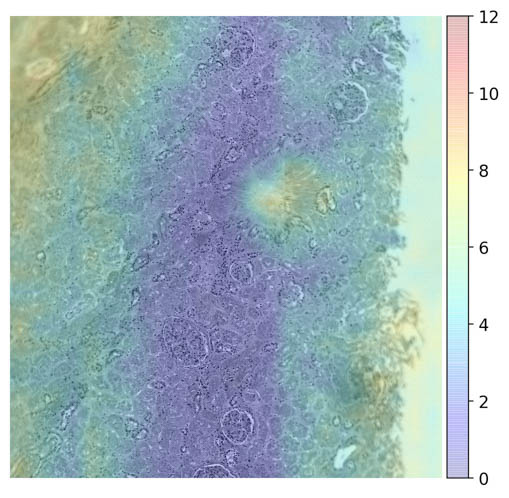}}
	
	\subfloat[][\centering{\scriptsize{2-kernel}}]{\includegraphics[width=0.2\textwidth]{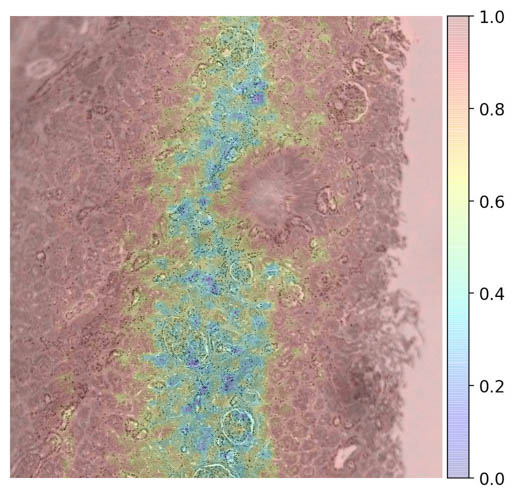}}
	\subfloat[][\centering{\scriptsize{10-kernel}}]{\includegraphics[width=0.2\textwidth]{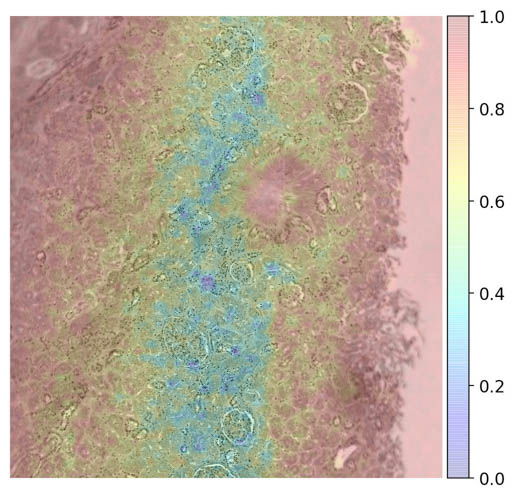}}
	\subfloat[][\centering{\scriptsize{DenseNet-13}}]{\includegraphics[width=0.2\textwidth]{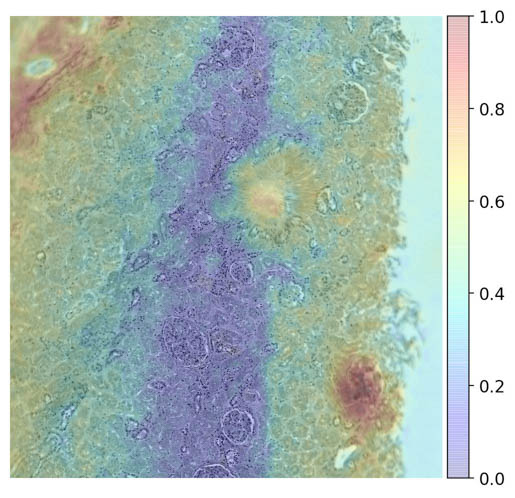}}
	\subfloat[][\centering{\scriptsize{ResNet-10}}]{\includegraphics[width=0.2\textwidth]{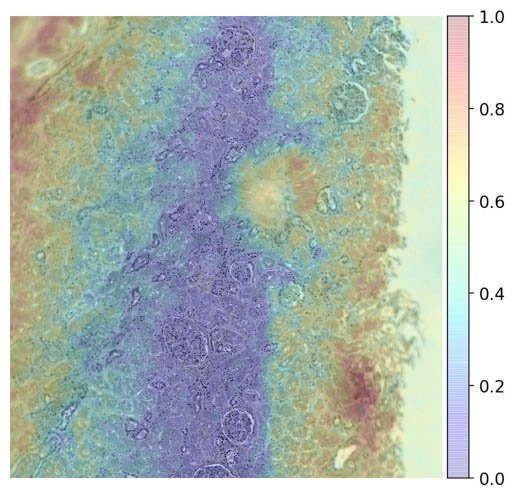}}\\
	
	\subfloat[][\centering{\scriptsize{ResNet-101}}]{\includegraphics[width=0.2\textwidth]{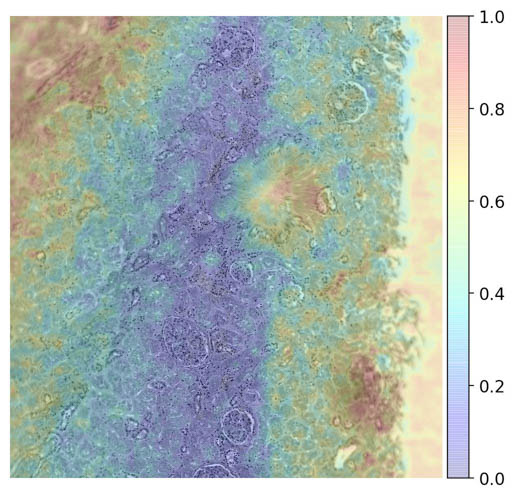}}
	\subfloat[][\centering{\scriptsize{FQPath}}]{\includegraphics[width=0.2\textwidth]{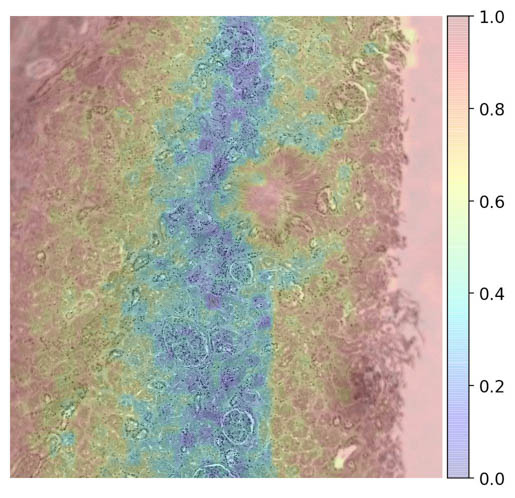}}
	\subfloat[][\centering{\scriptsize{HVS-MaxPol-l}}]{\includegraphics[width=0.2\textwidth]{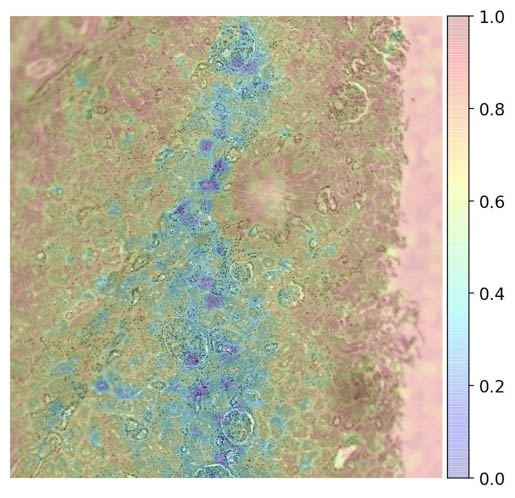}}
	\subfloat[][\centering{\scriptsize{HVS-MaxPol-2}}]{\includegraphics[width=0.2\textwidth]{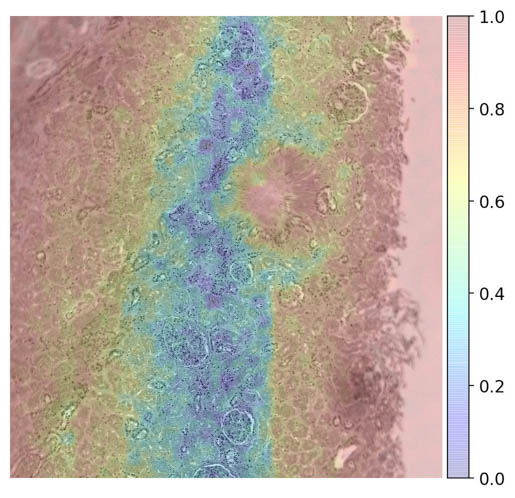}}\\
	
	\subfloat[][\centering{\scriptsize{Synthetic-MaxPol}}]{\includegraphics[width=0.2\textwidth]{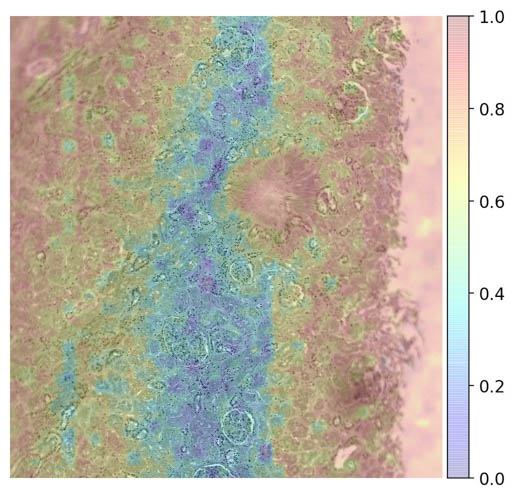}}
	\subfloat[][\centering{\scriptsize{LPC}}]{\includegraphics[width=0.2\textwidth]{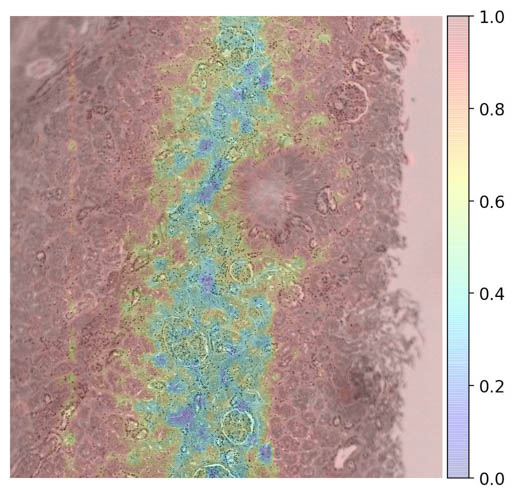}}
	\subfloat[][\centering{\scriptsize{GPC}}]{\includegraphics[width=0.2\textwidth]{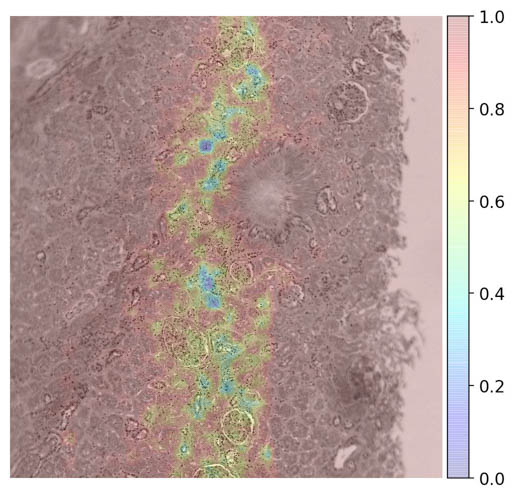}}
	\subfloat[][\centering{\scriptsize{SPRISH}}]{\includegraphics[width=0.2\textwidth]{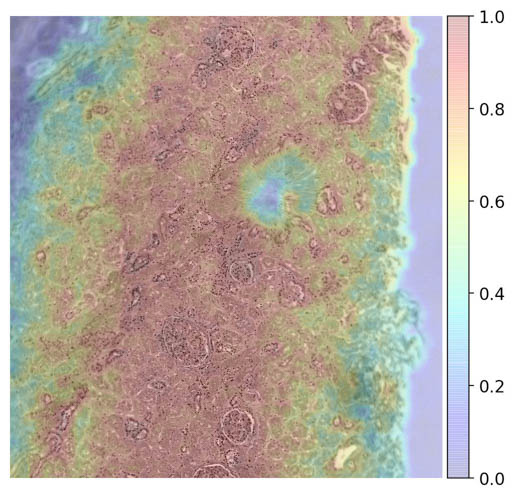}}\\
	
	\caption{Heat maps visualizations, higher score indicate more blurriness. \textbf{First row}: sample heat maps of a slide scanned at 40X of 3 models trained using MSE loss. The predicted scores correspond to absolute z-levels in the FocusPath dataset, ranging from 0 to 12.  \textbf{Second to fourth row}: normalized heat maps. The predicted scores of each model are independently linearly normalized to the range 0 to 1.}
	\label{heatmap_sup}
\end{figure}

\begin{figure}[h!]
	\centering
	\subfloat[][\centering{\scriptsize{2-kernel}}]{\includegraphics[width=0.2\textwidth]{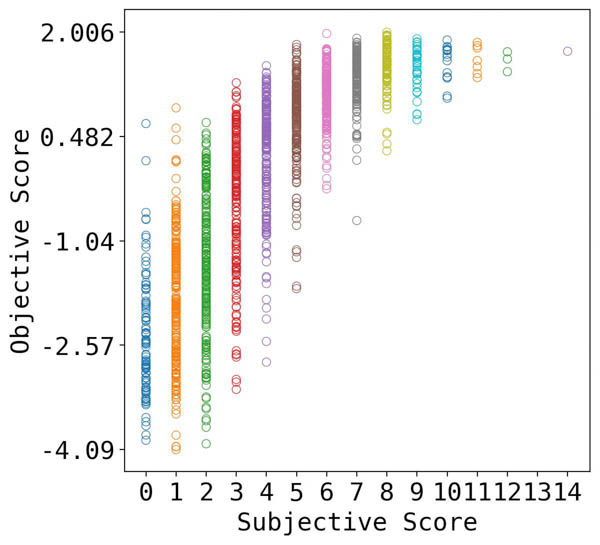}}
	\subfloat[][\centering{\scriptsize{10-kernel}}]{\includegraphics[width=0.2\textwidth]{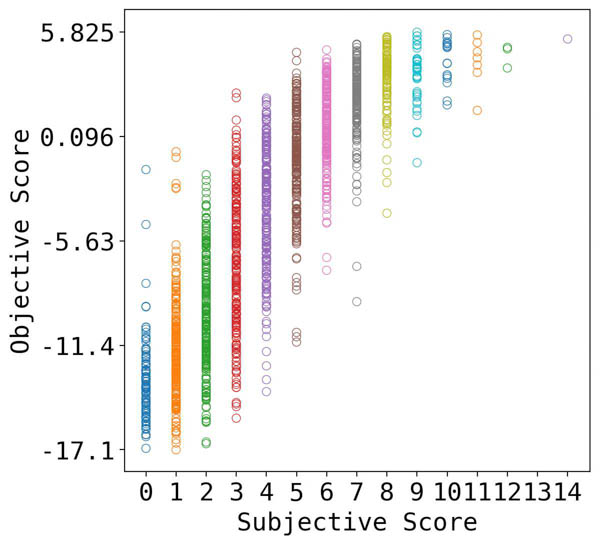}}
	\subfloat[][\centering{\scriptsize{DenseNet-13}}]{\includegraphics[width=0.2\textwidth]{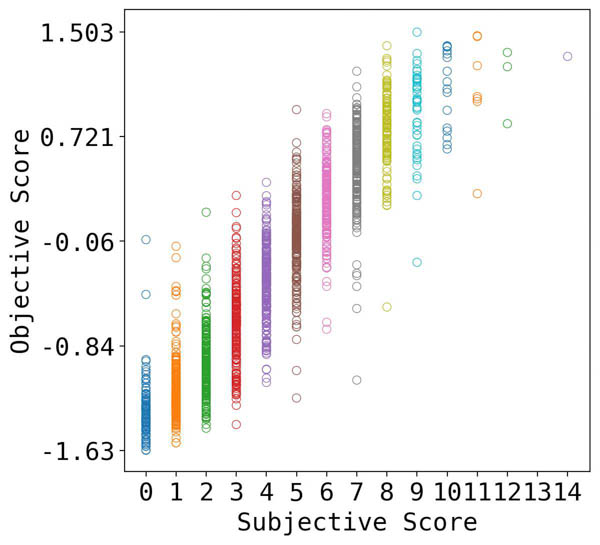}}
	\subfloat[][\centering{\scriptsize{ResNet-10}}]{\includegraphics[width=0.2\textwidth]{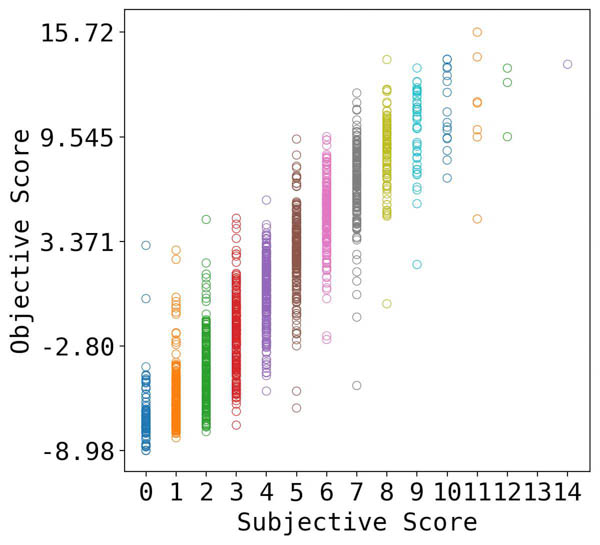}}\\
	\subfloat[][\centering{\scriptsize{ResNet-101}}]{\includegraphics[width=0.2\textwidth]{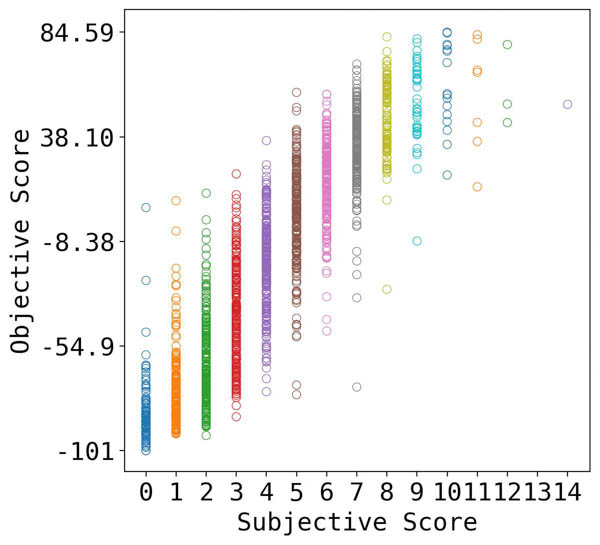}}
	\subfloat[][\centering{\scriptsize{FQPath}}]{\includegraphics[width=0.2\textwidth]{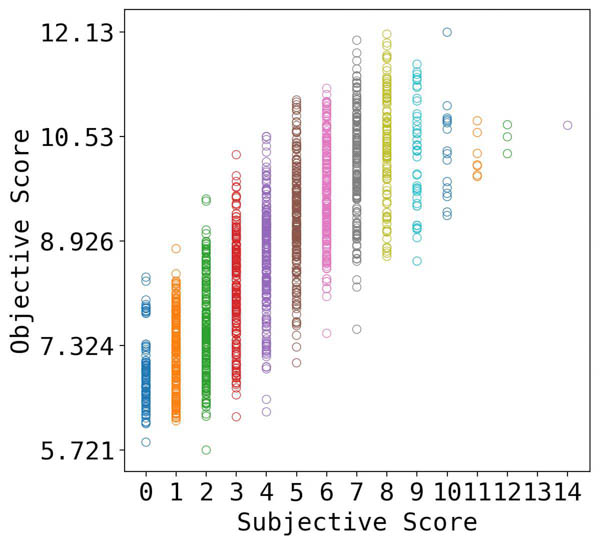}}
	\subfloat[][\centering{\scriptsize{Synthetic MaxPol}}]{\includegraphics[width=0.2\textwidth]{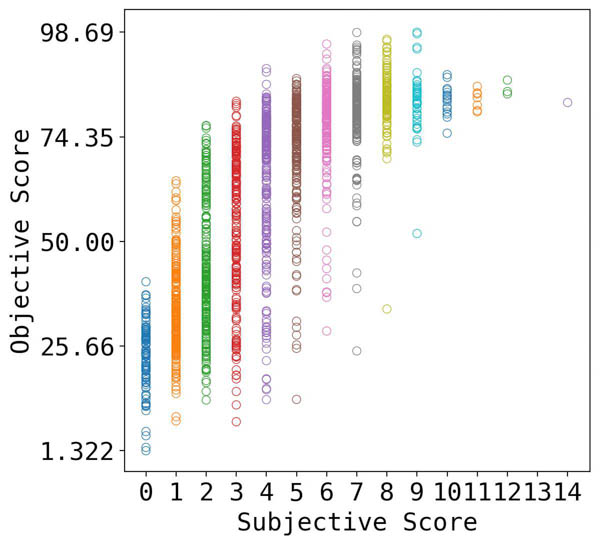}}
	\subfloat[][\centering{\scriptsize{LPC}}]{\includegraphics[width=0.2\textwidth]{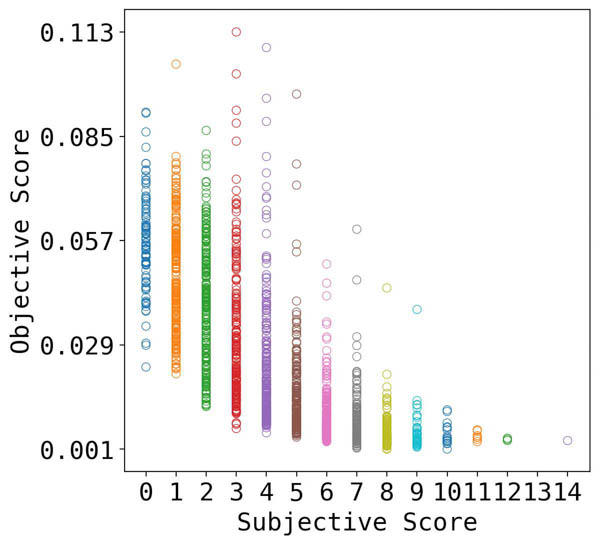}}\\
	\subfloat[][\centering{\scriptsize{GPC}}]{\includegraphics[width=0.2\textwidth]{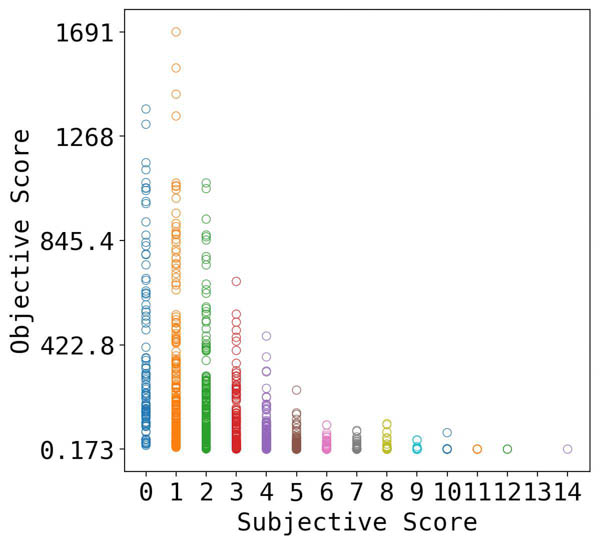}}
	\subfloat[][\centering{\scriptsize{SPARISH}}]{\includegraphics[width=0.2\textwidth]{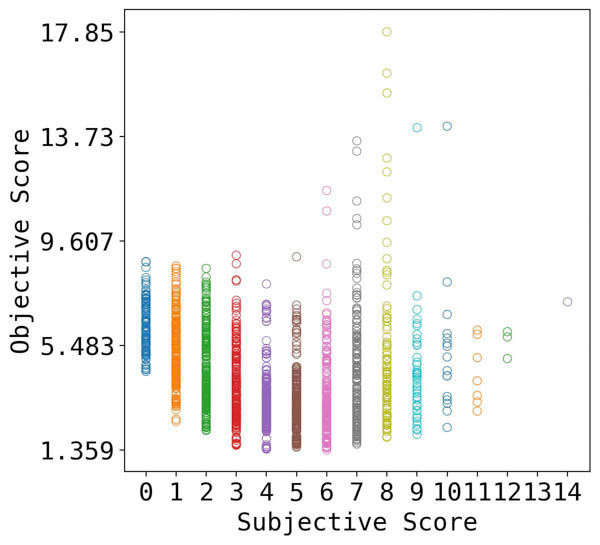}}
	\subfloat[][\centering{\scriptsize{HVS-MaxPol-1 }}]{\includegraphics[width=0.2\textwidth]{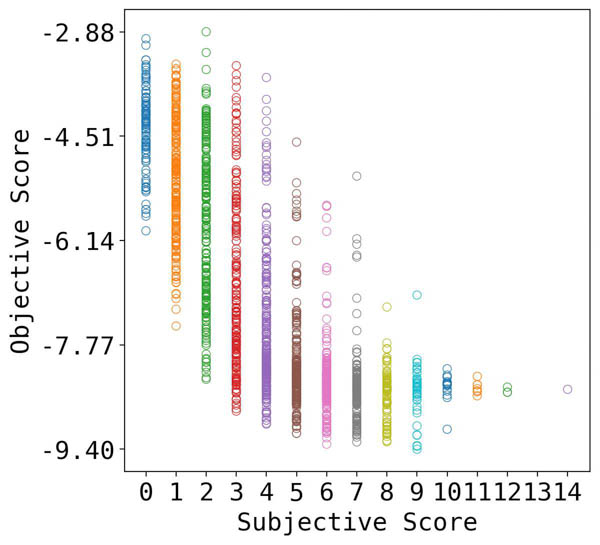}}
	\subfloat[][\centering{\scriptsize{HVS-MaxPol-2 }}]{\includegraphics[width=0.2\textwidth]{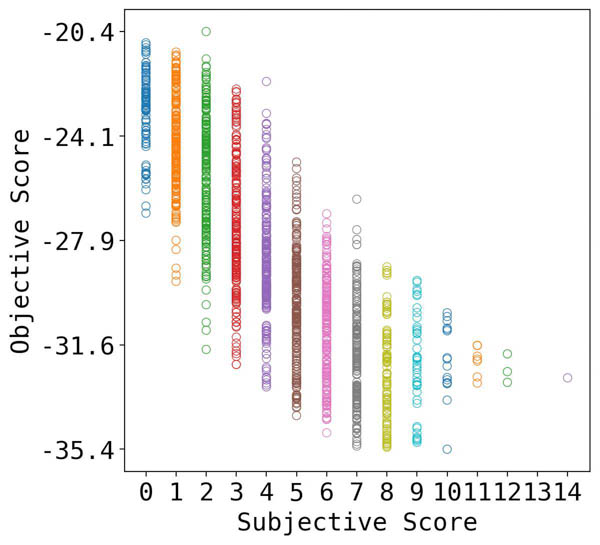}}\\
	\subfloat[][\centering{\scriptsize{2-kernel}}]{\includegraphics[width=0.2\textwidth]{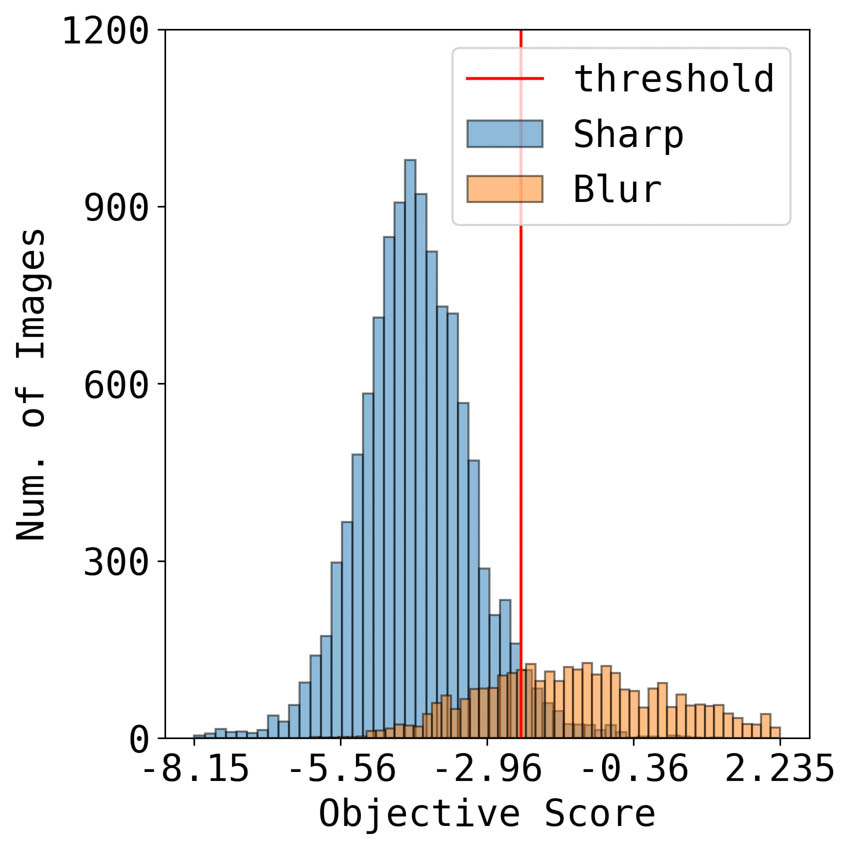}}
	\subfloat[][\centering{\scriptsize{10-kernel}}]{\includegraphics[width=0.2\textwidth]{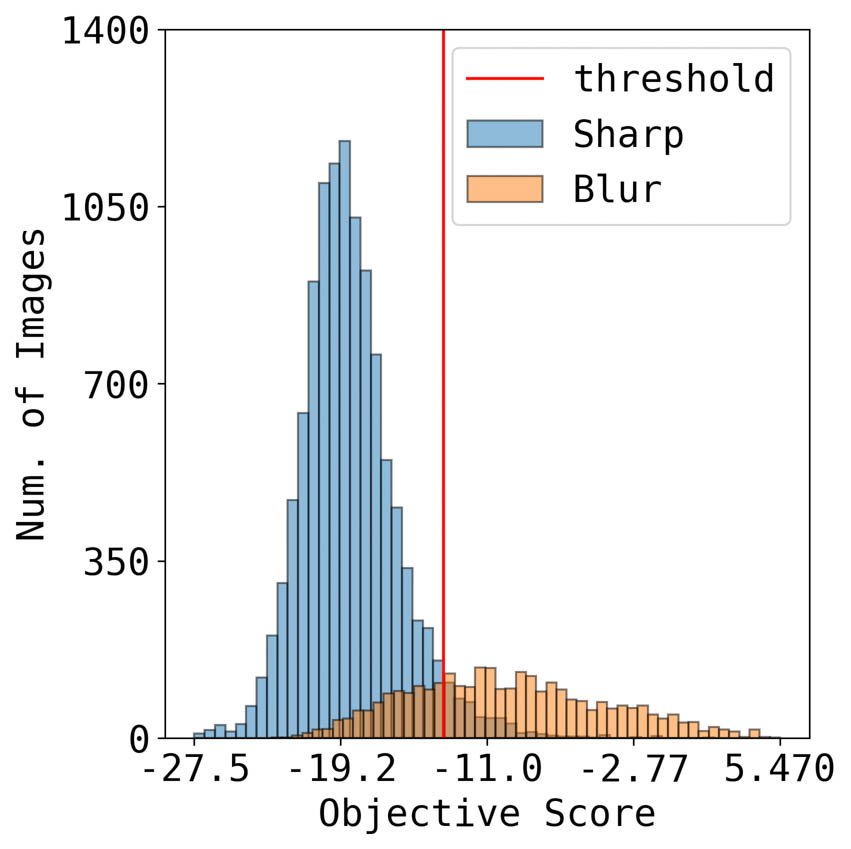}}
	\subfloat[][\centering{\scriptsize{DenseNet-13}}]{\includegraphics[width=0.2\textwidth]{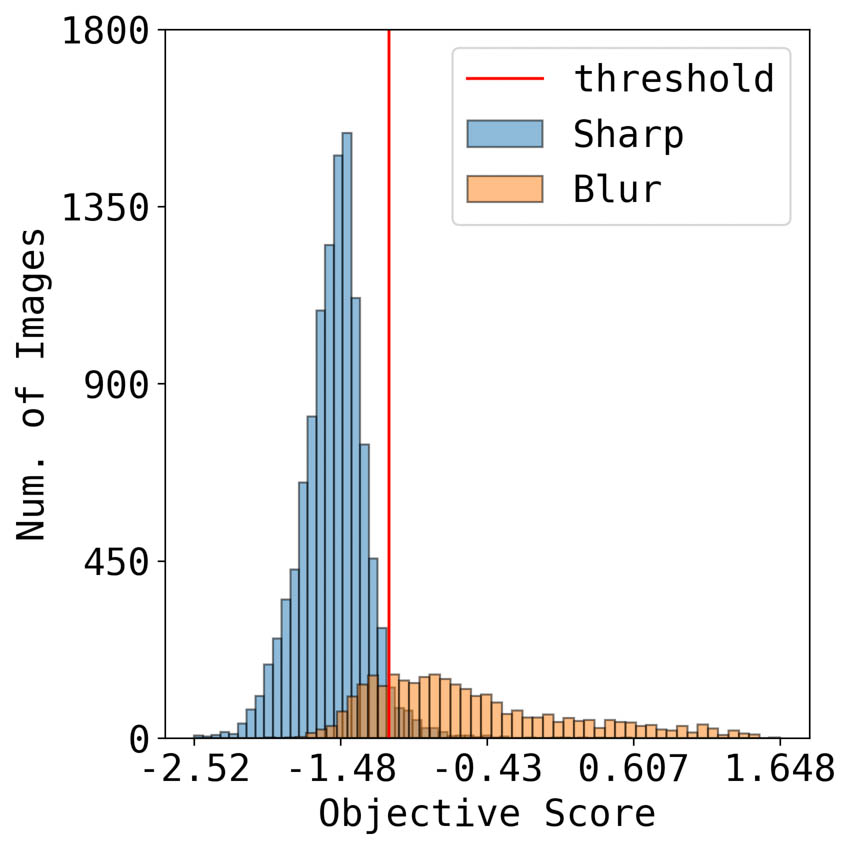}}
	\subfloat[][\centering{\scriptsize{ResNet-10}}]{\includegraphics[width=0.2\textwidth]{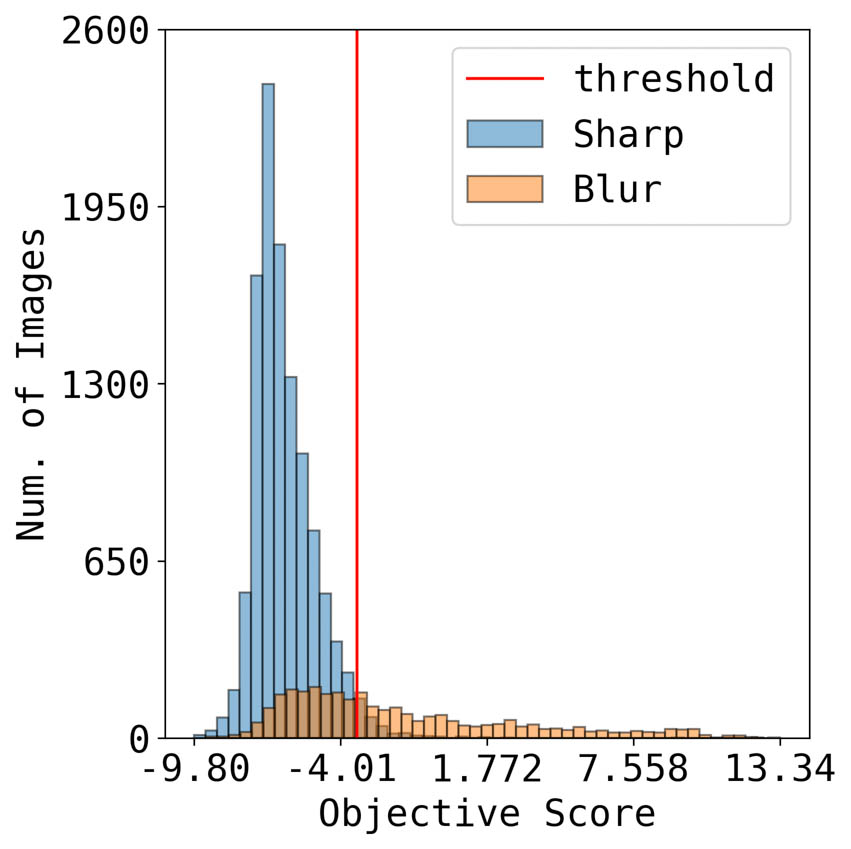}}\\
	\subfloat[][\centering{\scriptsize{ResNet-101}}]{\includegraphics[width=0.2\textwidth]{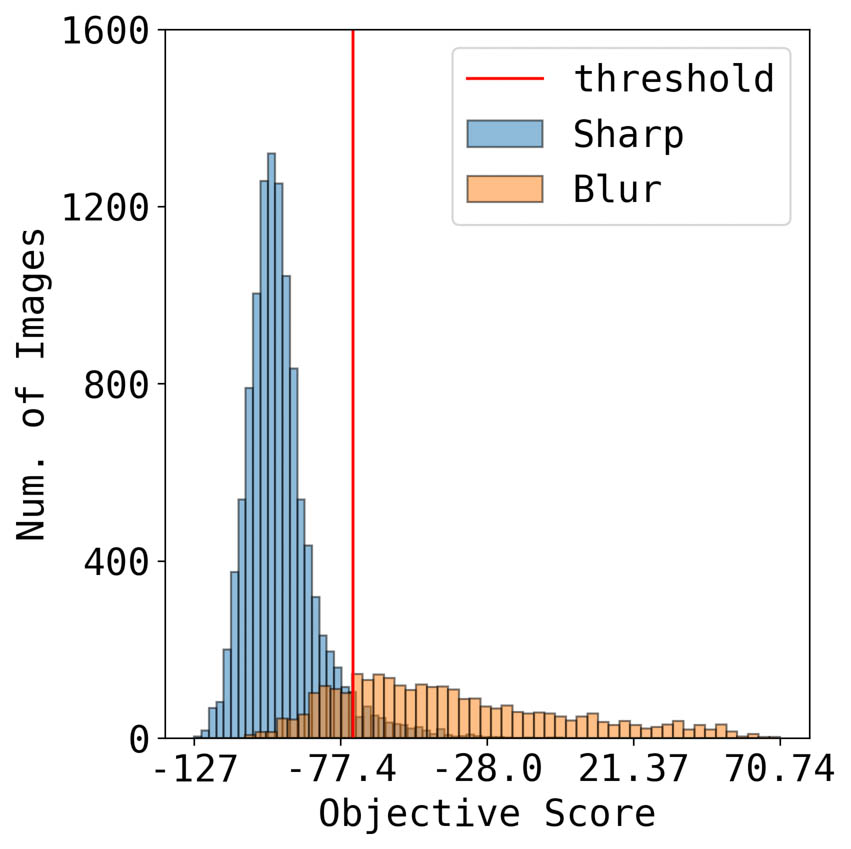}}
	\subfloat[][\centering{\scriptsize{FQPath}}]{\includegraphics[width=0.2\textwidth]{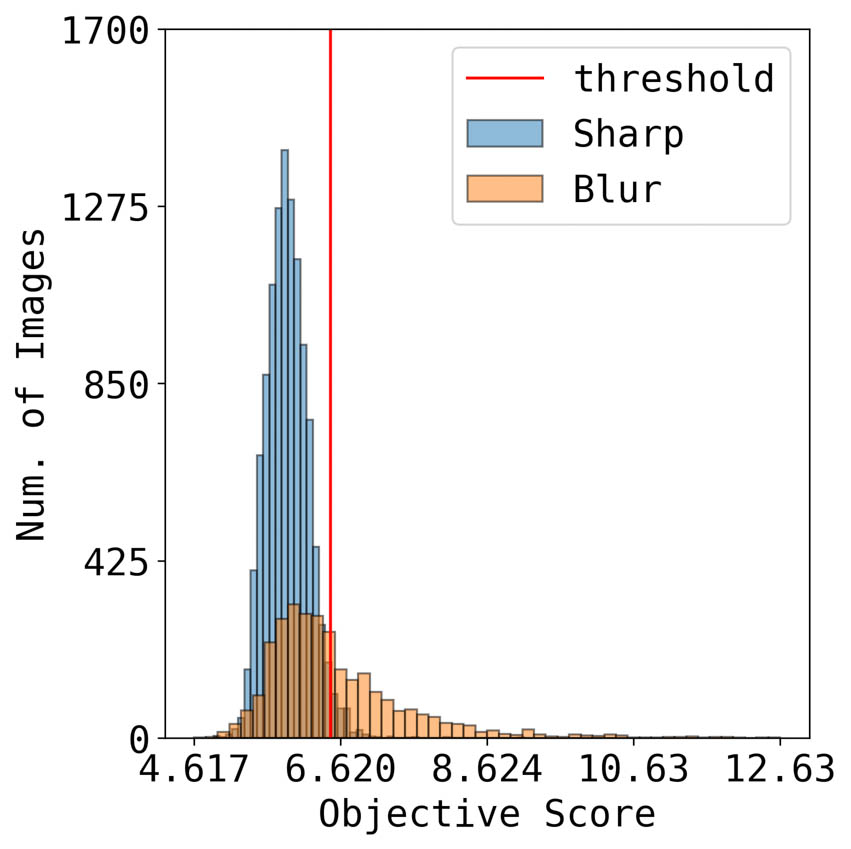}}
	\subfloat[][\centering{\scriptsize{Synthetic MaxPol}}]{\includegraphics[width=0.2\textwidth]{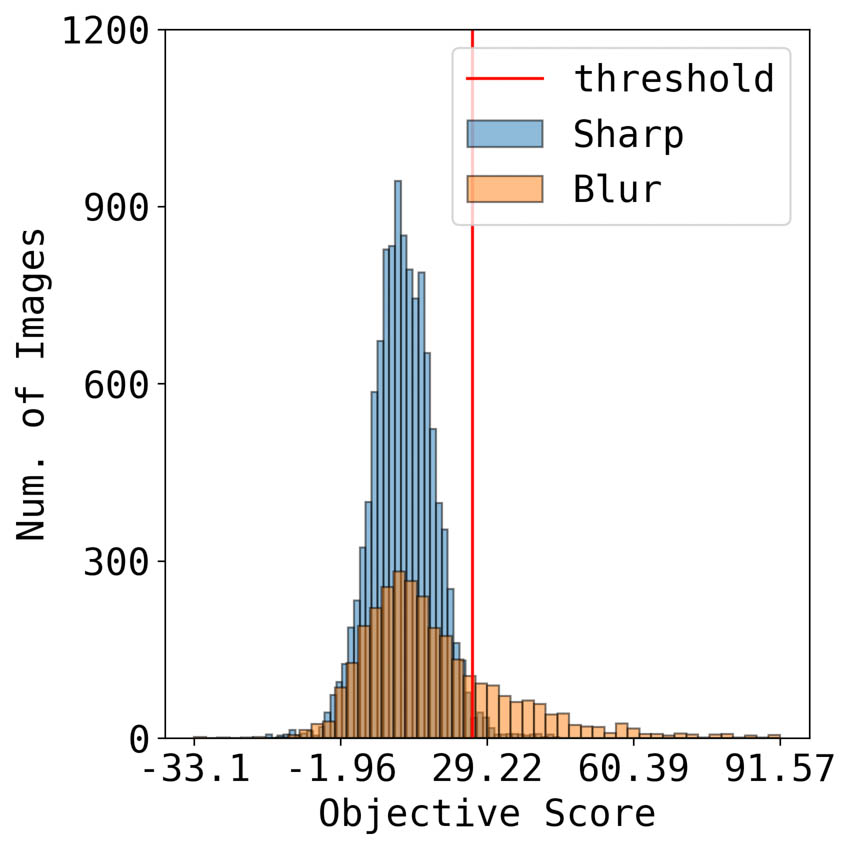}}
	\subfloat[][\centering{\scriptsize{LPC}}]{\includegraphics[width=0.2\textwidth]{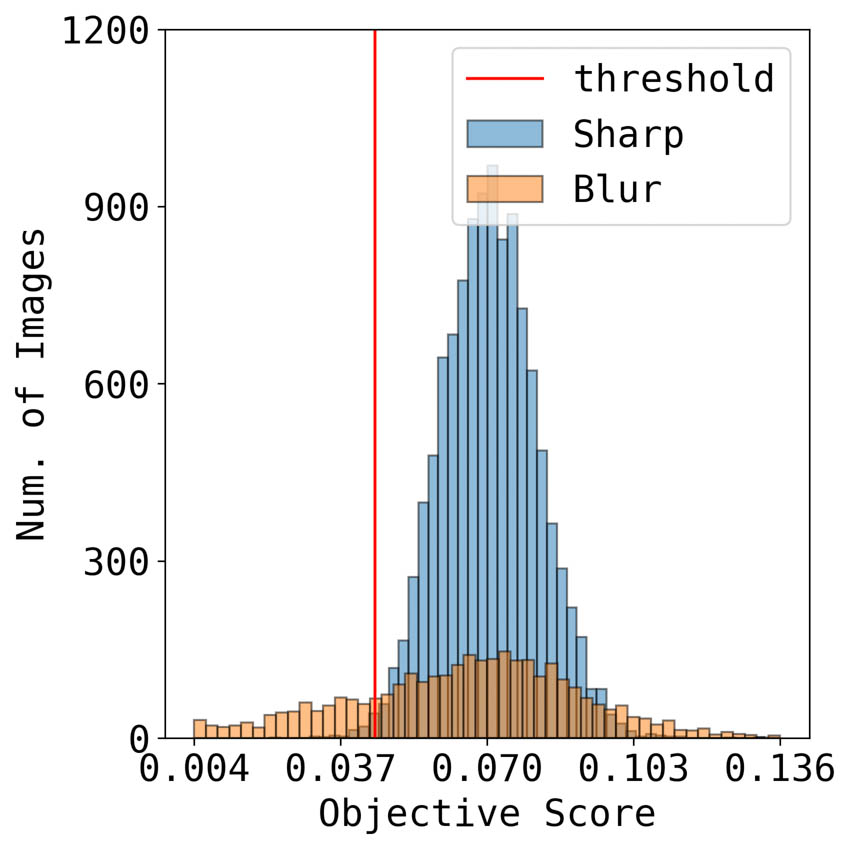}}\\
	\subfloat[][\centering{\scriptsize{GPC}}]{\includegraphics[width=0.2\textwidth]{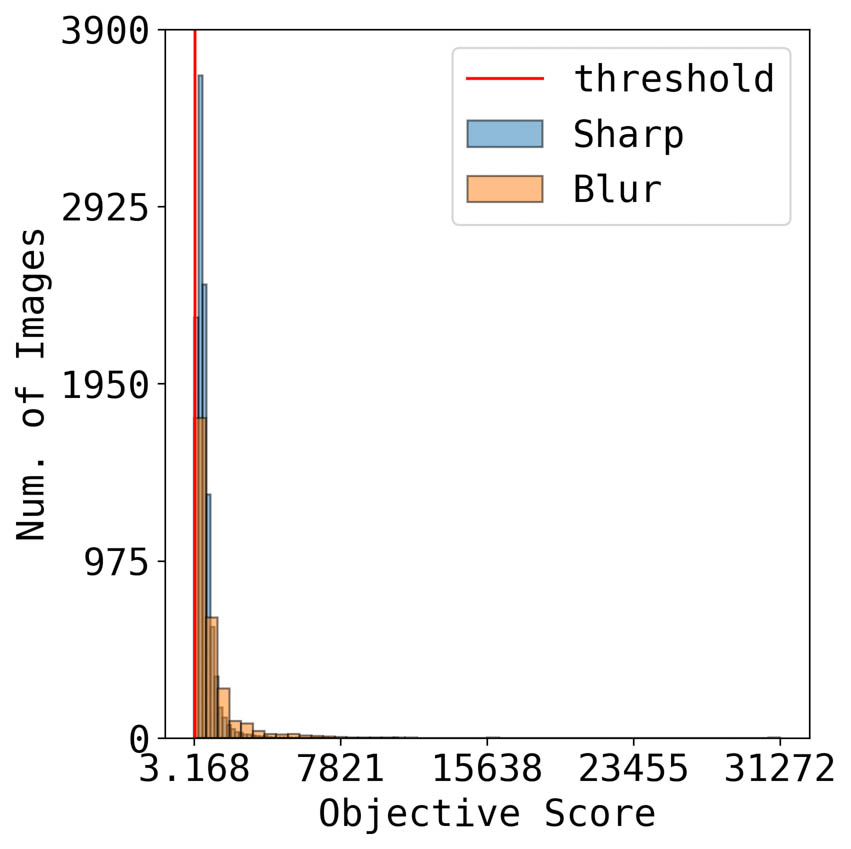}}
	\subfloat[][\centering{\scriptsize{SPARISH}}]{\includegraphics[width=0.2\textwidth]{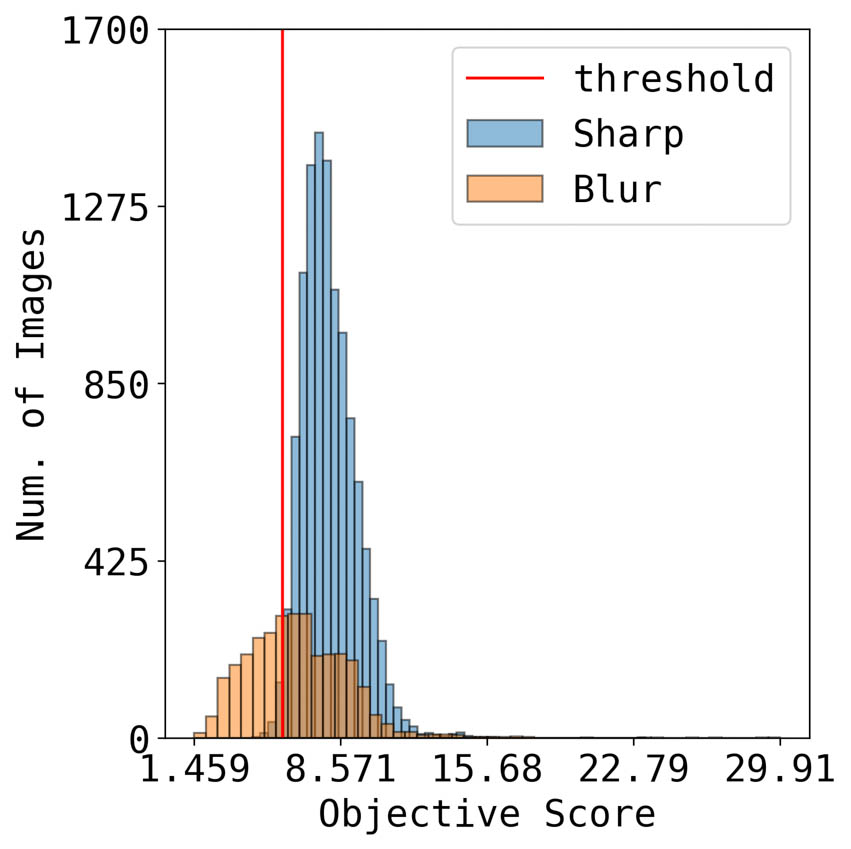}}
	\subfloat[][\centering{\scriptsize{HVS-MaxPol-1}}]{\includegraphics[width=0.2\textwidth]{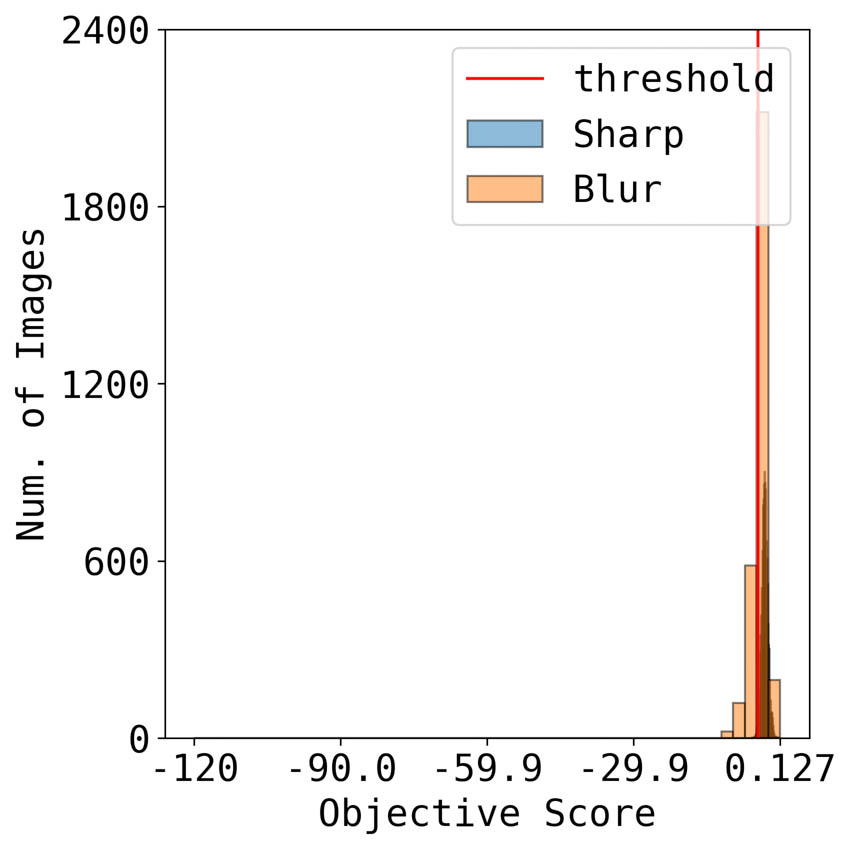}}
	\subfloat[][\centering{\scriptsize{HVS-MaxPol-2}}]{\includegraphics[width=0.2\textwidth]{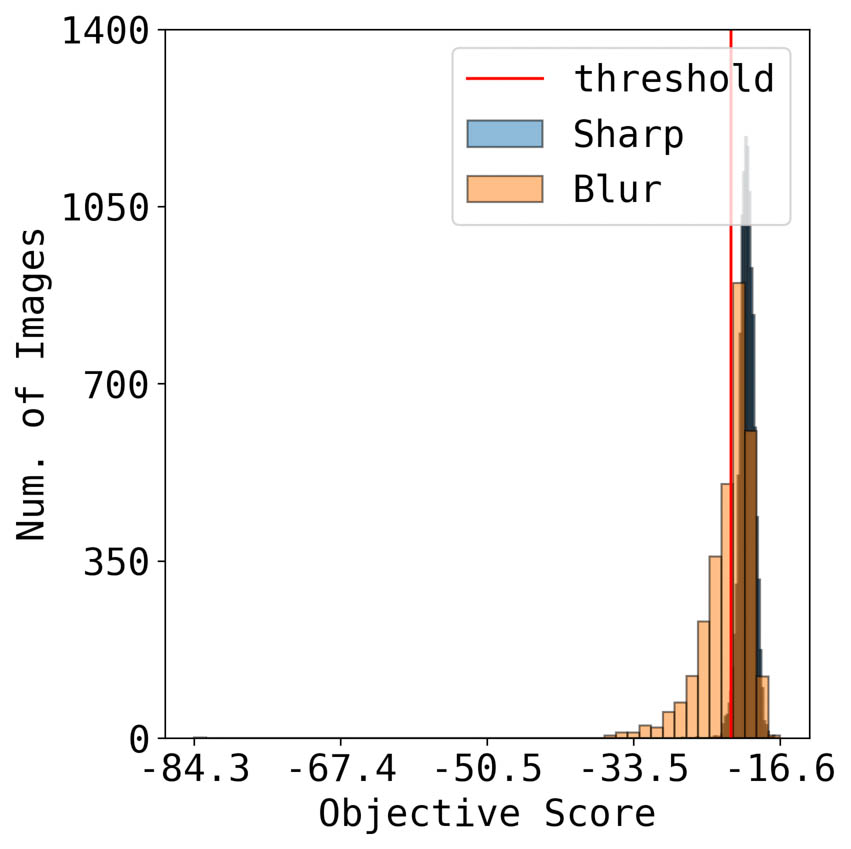}}\\
	\caption{Supplementary materials of Fig. $3$ in the main draft.}
	\label{eval_fig_sup}
\end{figure}

Figure \ref{heatmap_sup} demonstrates the complete heatmap visualization of all models and Figure \ref{eval_fig_sup} demonstrates complete evaluation results i.e. scatter histogram plots of all models.

\end{document}